\documentclass[letterpaper, preprint, paper,10pt]{AAS}	

\usepackage{bm}
\usepackage{amsmath}
\usepackage{amssymb}
\usepackage{subfigure}
\usepackage[colorlinks=true, pdfstartview=FitV, linkcolor=black, citecolor= black, urlcolor= black]{hyperref}
\usepackage{overcite}
\usepackage{footnpag}			      	
\usepackage{booktabs}
\usepackage[hang,flushmargin]{footmisc}

\PaperNumber{23-397}

\begin{document}

\title{4th body-induced secondary resonance overlapping inside unstable resonant orbit families: a Jupiter-Ganymede 4:3 + Europa case study}

\author{Bhanu Kumar\thanks{NSF Fellow, Jet Propulsion Laboratory, California Institute of Technology, 4800 Oak Grove Drive, Pasadena, CA 91109.},  
Rodney L. Anderson\thanks{Technologist, Jet Propulsion Laboratory, California Institute of Technology, 4800 Oak Grove Drive, Pasadena, CA 91109.},
\ and Rafael de la Llave\thanks{Professor, School of Mathematics, Georgia Institute of Technology, 686 Cherry St. NW, Atlanta, GA 30332.}
}

\maketitle{}

\begin{abstract}
The overlapping of mean-motion resonances is useful for low or zero-propellant space mission design, but while most related prior work uses a planar CRTBP model, tours of multi-moon systems require using resonances affected by two moons. In this case study, we investigate Jupiter-Ganymede unstable 4:3 resonant orbits in a concentric circular restricted 4-body Jupiter-Europa-Ganymede model. We show that despite their high order, secondary resonances between the 4:3 orbits and Europa have a large effect,  including 11/34, 12/37, 23/71, and 25/77. Computing newly generated objects inside the secondary resonances definitively confirms their overlap, which causes a complete structural change of the higher-energy unstable 4:3 orbits whose manifolds are most useful for low-TOF orbit transfers. We believe this phenomenon is general, with major implications for resonant orbit use in tour design.\end{abstract}

\section{Introduction}

\footnotetext{\copyright \, 2023. All rights reserved.}

In the planar circular restricted 3-body problem (PCRTBP), each mean motion resonance contains families of stable and unstable resonant periodic orbits across a range of energy levels. The unstable ones in particular are of special interest due to their attached stable/unstable manifolds; indeed, owing to the Chirikov resonance-overlap criterion \cite{chirikov1960}, it is the intersection of manifolds from different resonances that generates global chaos and enables large-scale natural transport across the system phase space. This instability in turn can be profitably leveraged for low or zero-propellant space mission trajectory design in multi-body systems, including multi-moon tours. For example, Anderson and Lo \cite{Anderson2010, Anderson2011, Anderson2011b} designed trajectories for hypothetical Europa missions using heteroclinics between Jupiter-Europa unstable resonant orbits. Vaquero and Howell \cite{vaqueroThesis} conducted a similar investigation in the Saturn system, using Saturn-Titan PCRTBP resonant periodic orbit manifolds for a theoretical mission to Hyperion. And more recently, the endgame mission design for the Europa Lander mission concept leveraged similar phenomena in its approach to the surface of Europa \cite{Anderson2021b}. See Anderson, Campagnola, and Lantoine \cite{Anderson2016} for examples of many other PCRTBP-based applications of resonant orbits.

Given the fundamental importance of unstable mean motion resonant orbits for tour design, an understanding of their properties is critical. In most prior work, the model used is the PCRTBP, which takes only the gravitation of one moon into account. Thus, the search for connections was carried out between orbits which were all resonant with the same moon. However, when designing tours of multi-moon systems, one must transition from orbits resonant with one moon to those resonant with a different moon; at some point, this necessitates finding a ``switching orbit'' \cite{KoLoMaRo} which allows the spacecraft to transition from resonances of one moon to those of another. The resonances which could play such a switching role will be contained in phase space regions which are significantly affected by both moons' gravity. Thus, at least a restricted 4-body model (R4BP) is needed to study these resonant orbits and any resulting transfers accurately. One such model, which we will use in this study, is the concentric circular restricted 4-body problem\cite{blazevski2012} (CCR4BP), where a third large body $m_{3}$, revolving in a circle around the largest mass $m_{1}$, is added to the PCRTBP. 

In this paper, we will demonstrate the importance of using an R4BP when studying unstable resonant orbits in such phase space regions by showing how the addition of $m_{3}$ can completely change the fundamental structure of the orbits being studied, compared to the PCRTBP. This will be accomplished through a case study on the effect of Europa on the unstable Jupiter-Ganymede 4:3 resonant orbit family. The dynamical mechanism by which this structural change occurs is the generation of secondary resonances between the periods of the PCRTBP unstable resonant periodic orbits and the synodic period of the forcing from $m_{3}$, which can undergo Chirikov overlapping \emph{inside} the unstable resonant orbit family if the forcing from $m_{3}$ is strong enough; we find this to indeed be the case for the Jupiter-Europa-Ganymede 4:3 orbits. As secondary resonances create entirely new types of orbits, their overlap has major implications for the kinds of resonant orbits which should be analyzed in such multi-moon systems. Furthermore, these new orbits take the place of precisely the higher-energy, closer moon-flyby, more highly-unstable resonant orbits whose stable/unstable manifolds are the most useful ones for practical, lower time-of-flight (TOF) trajectories. While secondary resonance overlapping has been studied in the context of families of stable, maximal-dimensional librational tori inside mean motion\cite{batyginMorby, paez2018, pichierriMorby} or spin-orbit\cite{gkoliasEtAl} resonances in celestial systems, to our knowledge this study is the first demonstration of such phenomena for unstable, non-maximal dimensional mean motion resonant orbits.

In this paper, we start with a summary of the CCR4BP model and its stroboscopic map, followed by an overview of some of the mathematical concepts which will be needed, namely normally hyperbolic invariant manifolds and Chirikov resonance overlapping. We next discuss the computational tools used in this study for computing various types of invariant objects. After reviewing some related preliminary observations from our prior study\cite{kumar2023}, we then present the methodology and results of a non-resonant torus-continuation based investigation of the growth of Europa-induced secondary resonances inside the 4:3 Jupiter-Ganymede unstable orbit family. Finding very strong evidence of these secondary resonances growing large enough to overlap, we next directly compute the most important dynamical structures contained inside the secondary resonances themselves in the full Jupiter-Europa-Ganymede CCR4BP, which yields definitive confirmation of the overlap. We conclude with a discussion on the implications of this secondary resonance overlap for studies involving unstable mean motion resonances in multi-moon systems, as well as for tour design. 

\section{Planar Concentric Circular Restricted 4BP} \label{modelsection}
The planar concentric circular restricted 4-body problem \cite{blazevski2012} (CCR4BP) describes the motion of a spacecraft influenced by the gravity of three large masses $m_{1}$, $m_{2}$, and $m_{3}$ with $m_{1} >> m_{2}, m_{3}$. $m_{2}$ and $m_{3}$ are assumed to revolve around $m_{1}$ in coplanar, concentric circles of radii $r_{12}$ and $r_{13}$, where $m_{2}$ has no effect on the motion of $m_{3}$ nor vice versa. Indeed, taking $m_{1}$, $m_{2}$, and $m_{3}$ to be Jupiter, Ganymede, and Europa, the motion of $m_{1}$, $m_{2}$, and $m_{3}$ just described does not satisfy the full 3-body problem, but nevertheless approximates the true physical system very well. Due to Kepler's third law, the angular velocities $\Omega_{2}$ and $\Omega_{3}$ of the revolution of $m_{2}$ and $m_{3}$ around $m_{1}$ depend on the masses and the orbital radii, as $\Omega_{i} = \sqrt{{\mathcal{G}(m_{1}+m_{i})}{r_{1i}^{-3}}}$ for $i = 2,3$ where $\mathcal{G}$ denotes the universal gravitational constant. In the planar CCR4BP, the circular orbits of $m_{2}$ and $m_{3}$ as well as the spacecraft trajectory are assumed to all lie in the same plane. 

Now, let $\mu = \frac{m_{2}}{m_{1}+ m_{2}}$ and  $\mu_{3} = \frac{m_{3}}{m_{1}+ m_{2}}$ be the mass ratios. As in the PCRTBP, one can normalize mass, length, and time units so that $\mathcal{G}(m_{1}+m_{2})$, $r_{12}$, and $\Omega_{2}$ are all $1$. Then, the planar CCR4BP equations of motion can be written in the same synodic coordinate system usually used for the CRTBP; $m_{1}$ and $m_{2}$ lie on the synodic frame $x$-axis, and the $m_{1}$-$m_{2}$ barycenter is taken as the frame origin. With these units and coordinate frame, the angle between the position of $m_{3}$ and the $x$-axis at time $t$ will be $\theta_{3}(t) = (\Omega_{3}-1)t + \theta_{3,0}$; the position of $m_{3}$ thus is $(x_{3}(t), y_{3}(t)) = (-\mu+r_{13} \cos(\theta_{3}), r_{13} \sin(\theta_{3}) )$. The equations of motion can then be written in position-momentum space $(x,y,p_{x},p_{y})$ as (see Blazevski and Ocampo \cite{blazevski2012} for a derivation)
\begin{equation} \label{ccr4bpEOM} \begin{aligned} \dot x= p_{x} + y \quad & \quad \dot y= p_{y} - x \\
 \dot p_x=  p_{y} - (1-\mu)\frac{x+\mu}{r_{1}^{3}} &- \mu \frac{x - (1-\mu)}{r_{2}^{3}} - \mu_{3} \frac{x - x_{3}}{r_{3}^{3}} - \mu_{3} \frac{\cos \theta_{3}}{r_{13}^{2}} \\
 \dot p_y=  -p_{x} - (1-\mu)\frac{y}{r_{1}^{3}} &- \mu \frac{y}{r_{2}^{3}} - \mu_{3} \frac{y - y_{3}}{r_{3}^{3}} - \mu_{3} \frac{\sin \theta_{3}}{r_{13}^{2}} \end{aligned} \end{equation}
where $r_{1} = \sqrt{(x+\mu)^{2} + y^{2}}$, $r_{2} = \sqrt{(x-1+\mu)^{2} + y^{2}}$, and $r_{3} = \sqrt{(x-x_{3})^{2} + (y-y_{3})^{2}}$ are the distances from the spacecraft to $m_{1}$, $m_{2}$, and $m_{3}$, respectively.  Note that when $\mu_{3}=0$, Eq. \eqref{ccr4bpEOM} is just the $m_{1}$-$m_{2}$ PCRTBP. The equations of motion Eq. \eqref{ccr4bpEOM} are Hamiltonian, with time-periodic Hamiltonian function
\begin{equation}  \label{ccr4bpH} \begin{aligned}  H_{\mu_{3}}(x,y,p_x,p_{y}, \theta_{3})= \frac{p_{x}^{2}+p_{y}^{2}}{2} + p_{x}y -p_{y}x  - \frac{1-\mu}{r_{1}} - \frac{\mu}{r_{2}}  - \frac{\mu_{3}}{r_{3}} + \mu_{3} \frac{x \cos \theta_{3}}{r_{13}^{2}} + \mu_{3} \frac{y \sin \theta_{3}}{r_{13}^{2}} \end{aligned}  \end{equation}

In Eq. \eqref{ccr4bpEOM}, the CCR4BP equations of motion are written in the synodic $m_{1}$-$m_{2}$  frame, with units defined to make $r_{12}=\mathcal{G}(m_{1}+m_{2})=\Omega_{2}=1$. One can also write the CCR4BP equations of motion in an $m_{1}$-$m_{3}$ synodic frame centered at the $m_{1}$-$m_{3}$ barycenter, however, with units normalized such that $r_{13}=\mathcal{G}(m_{1}+m_{3})=\Omega_{3}=1$. The $m_{1}$-$m_{3}$ frame equations of motion maintain the same form as Eq. \eqref{ccr4bpEOM}, but with all subscripts 2 and 3 swapped, new perturbation phase $\theta_{2}=-\theta_{3}$, and new mass ratios $\bar \mu = \frac{m_{3}}{m_{1}+ m_{3}}$ and  $\bar \mu_{2} = \frac{m_{2}}{m_{1}+ m_{3}}$ replacing $\mu$ and $\mu_{3}$, respectively; see Kumar et al \cite{kumar2021aug} for the full equations of motion in this case. Similar to the $m_{1}$-$m_{2}$ frame, setting $\bar \mu_{2}=0$ in this frame yields the $m_{1}$-$m_{3}$ PCRTBP equations of motion.

\subsection{Stroboscopic Maps} \label{stroboscopic}

The CCR4BP flow is defined on a 5D extended phase space, $(x,y, p_{x}, p_y, \theta_{3}) \in \mathbb{R}^{4} \times \mathbb{T}$ in the $m_{1}$-$m_{2}$ frame CCR4BP or similarly $(\bar x, \bar y,  \bar p_x, \bar p_y, \theta_{2}) \in \mathbb{R}^{4} \times \mathbb{T}$ in the $m_{1}$-$m_{3}$ frame. The perturbation phase angle $\theta_{3}$ or $\theta_{2}$ increases the phase space dimension by 1 compared to the 4D phase space of the PCRTBP $(x,y, p_{x}, p_y)$. However, one can instead use a stroboscopic map which reduces the dimension of the CCR4BP back to 4D, which has a number of benefits both for computations as well as for comparisons with the PCRTBP. 

Let $p=3$ or 2 be such that $\theta_{p}$ is the perturbation phase angle of the frame being used. Now, define the stroboscopic map $F: \mathbb{R}^{4} \times \mathbb{T} \rightarrow \mathbb{R}^{4} \times \mathbb{T}$ as the time-$\frac{2\pi}{|\Omega_p-1|}$ mapping of extended phase space points by the CCR4BP equations of motion. Since $\dot \theta_{p} = \Omega_{p}-1$, in time-$\frac{2\pi}{|\Omega_p-1|}$, the angle $\theta_{p}$ will revolve by exactly $2\pi$. Thus, the $\theta_{p}$ component of $F(x,y,p_{x},p_{y}, \theta_{p,f}) $ will simply be $\theta_{p,f}$ again, for any $(x,y,p_{x},p_{y}) \in \mathbb{R}^{4}$. As the $\theta_{p}$ state component is $F$-invariant, one can fix its value to any $\theta_{p,f}$ (we take $\theta_{p,f}=0$) and consider the dynamics of $F$ on the 4D subspace $(x,y,p_{x},p_{y}; \theta_{p,f})$ instead of the CCR4BP flow on its 5D phase space. 

The stroboscopic map preserves all the dynamical properties of the CCR4BP, including invariant sets and manifolds; each CCR4BP orbit has a stroboscopic map counterpart and vice versa. From this point onwards, we make a slight abuse of notation and consider $F:\mathbb{R}^{4} \rightarrow \mathbb{R}^{4}$. Usually we will use a subscript, such as $F_{\mu_{3}}$, to signify the dependence of $F$ on the CCR4BP mass parameter. Note that the stroboscopic map definition is valid even for $\mu_{3}=0$ or for $\bar \mu_{2}=0$, which are just the $m_{1}$-$m_{2}$ and $m_{1}$-$m_{3}$ PCRTBP respectively.
 
\section{Normally Hyperbolic Invariant Manifolds}

The PCRTBP contains many families of unstable periodic orbits, including planar Lyapunov orbits at L1 and L2 as well as unstable resonant periodic orbits. Each of these unstable orbit families can be parameterized by a single parameter, oftentimes taken as the orbit Jacobi constant (or equivalently, its energy). Now, consider such an unstable periodic orbit family which has one orbit per value of energy $E$ over some range of energy values $[E_{min}, E_{max}]$. Each orbit in the family is topologically equivalent to a circle $\mathbb{T}$, so if one takes the set of all the periodic orbits over all values of $E \in [E_{min}, E_{max}]$, the resulting set $\Xi$ will be a 2D invariant manifold topologically equivalent to the cylinder $\mathbb{T} \times [0,1]$ in the 4D PCRTBP phase space. Moreover, as the unstable periodic orbits which foliate $\Xi$ have stable and unstable eigenvectors, at each point of $\Xi$ these eigenvectors will represent stable and unstable directions transverse to (i.e. not tangent to) the manifold $\Xi$. 

Now, at any point of $\Xi$, taking the 2D tangent vector space of $\Xi$ together with the stable and unstable directions at that point gives a set of vectors spanning the entire 4D phase space. Thus, each point of $\Xi$ has a vector basis for the entire phase space where the effect of the PCRTBP state transition matrix causes all basis vectors transverse to $\Xi$ to exponentially contract (the stable direction) or expand (the unstable direction) over time, at rates much stronger than those of any such contraction/expansion in the basis directions tangent to $\Xi$. This is the fundamental property which defines a \emph{normally hyperbolic invariant manifold}, or NHIM for short. For a rigorous definition of NHIMs, see Fenichel\cite{fenichel1971persistence}. Note that $\Xi$ as defined above is not only a NHIM for the PCRTBP flow, but also for the corresponding stroboscopic map $F_{\mu_{3}=0}$ defined in the previous section. 

The key property of NHIMs which will be important is that they persist under sufficiently small perturbations of the original dynamical system into the new system\cite{fenichel1971persistence, hirschPughShub}. In fact, as long as the previously mentioned conditions on rates of transverse versus tangent expansion/contraction hold, then the size of the perturbation under which the NHIM persists can even be rather large in practice. The perturbed NHIM may deform in phase space, but it will remain topologically equivalent to the NHIM from the unperturbed system. Thus, the cylindrical 2D NHIM $\Xi_{\mu_{3}=0}=\Xi$ formed by any PCRTBP unstable periodic orbit family can be expected to persist into the CCR4BP stroboscopic map $F_{\mu_{3}}$ system when $\mu_{3}>0$ as well, in the form of a cylindrical 2D NHIM $\Xi_{\mu_{3}}$ (here it is key that the CCR4BP stroboscopic map is defined on the same phase space $\mathbb{R}^{4}$ as the PCRTBP map $F_{\mu_{3}=0}$, which is required in order to apply the persistence theorems for NHIMs). 

However, though the NHIM as a whole may persist into a perturbed system, there is no guarantee that the dynamical objects \emph{inside} the NHIM do the same. Indeed, the PCRTBP NHIM $\Xi_{\mu_{3}=0}$ was entirely foliated by flow-periodic orbits, which are invariant circles (1D tori) of the stroboscopic map $F_{\mu_{3}=0}$. These tori will have a range of rotation numbers $\omega$, some of which will be resonant. From KAM theory\cite{delshams2008scattering}, we know that for sufficiently small perturbations, tori at ``sufficiently irrational'' $\frac{\omega}{2\pi}$ will persist inside the CCR4BP NHIM $\Xi_{\mu_{3}}$, while resonant circles (at rational $\frac{\omega}{2\pi}$) will disappear as soon as $\mu_{3} > 0$. However, as will be seen in this paper, the persistence of tori predicted by KAM theory is less robust than the persistence of NHIMs; that is, a perturbation which is ``sufficiently small'' for NHIM persistence may not be small enough for the persistence of invariant tori, even at irrational $\omega$. Thus, the dynamics inside a NHIM may still change drastically with $\mu_{3}$.  

The above discussion was done for the NHIM $\Xi_{\mu_{3}=0}$ being any family of PCRTBP unstable periodic orbits, but for the remainder of this paper, we will let $\Xi_{\mu_{3}=0}$ represent the Jupiter-Ganymede PCRTBP 4:3 internal unstable resonant orbit family, with $\Xi_{\mu_{3}}$ being its persisting counterpart under the CCR4BP stroboscopic map. To analyze the dynamics inside the aforementioned NHIM $\Xi_{\mu_{3}}$, it is useful to recall that $\Xi_{\mu_{3}}$ is just a 2D cylinder. Thus, $F_{\mu_{3}}$ restricted to $\Xi_{\mu_{3}}$ is effectively a 2D symplectic\cite{delshams2008scattering} map from $\mathbb{T} \times [0,1]$ into itself. So, for visualization and understanding of the NHIM's internal dynamics, a 2D plot gives complete information. In addition, since for $\mu_{3}=0$ the dynamics inside $\Xi_{\mu_{3}}$ are comprised entirely of invariant circles, $F_{\mu_{3}}$ is not only symplectic, but also near-integrable inside $\Xi_{\mu_{3}}$ for $\mu_{3}$ ``small''. Thus, we can use known results for 2D symplectic, near-integrable maps on cylinders to characterize the dynamics inside the NHIM $\Xi_{\mu_{3}}$. One such result which will be of fundamental importance in this study is the Chirikov resonance overlap criterion. 

\section{Resonances and Chirikov's Overlap Criterion}

\begin{figure}
\includegraphics[width=0.33\columnwidth]{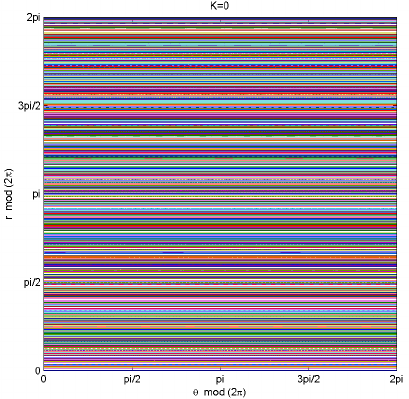}
\includegraphics[width=0.33\columnwidth]{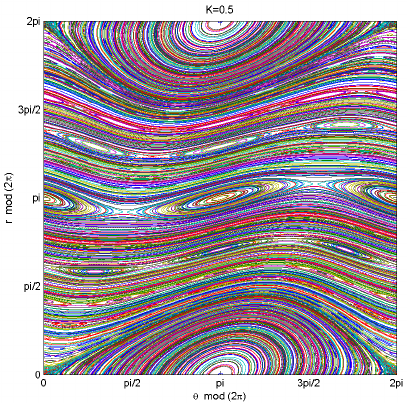}
\includegraphics[width=0.33\columnwidth]{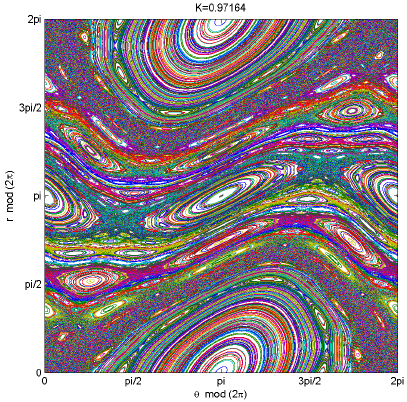}
\caption{ \label{fig::chirikov} Standard map phase space trajectories for $K=0, 0.5, 0.97164$ (left to right)\cite{standardMapPics}}
\vspace{-18pt}
\end{figure}

The Chirikov resonance overlap criterion, first introduced\cite{chirikov1960} in 1959, is a physical criterion for explaining the transition from regular motions to chaotic global instability in deterministic, near-integrable Hamiltonian and symplectic systems. While a fully general explanation is beyond the scope of this paper, an illustrative example can be seen in the Chirikov standard map, a  2D symplectic map defined on the space $(\theta, r) \in \mathbb{T} \times \mathbb{R}$ through the equations $r_{n+1} = r_{n} + K \sin(\theta_{n})$, $\theta_{n+1} = \theta_{n}+ r_{n+1}$. For $K=0$, $r$ is constant on each trajectory while $\theta$ simply rotates by $r$ each iteration; the entire phase space is foliated by invariant circles which ``circulate'', i.e. wrap across the entire range of $\theta$ values from 0 to $2\pi$, as is illustrated in the leftmost plot of Figure \ref{fig::chirikov}. Such a phase space structure, comprised entirely of tori, is called \emph{integrability}. 

For $K>0$ small enough, many of the circulating invariant circles present in the $K=0$ case persist under the perturbation, as is visible in the middle plot of Figure \ref{fig::chirikov}. However, not all these circles persist; it is precisely the circulating circles at rotation numbers $r=2\pi \frac{m}{n}$, $m, n \in \mathbb{Z}$ which disappear first. Such rotation numbers correspond to resonances; moreover, it is also visible in Figure \ref{fig::chirikov} that each resonance does not only affect the dynamics exactly at that $r$ value, but also in a region around it. Such resonant ``islands'' are clearly visible in the figure at the lower order resonances $r/2\pi=0/1, 1/3, 1/2, 2/3$, and $1/1$. The width of each island in the $r$ direction will be proportional to the square root of the perturbation $K$\cite{morbyBook}; thus, as $K$ grows, one can predict that eventually consecutive resonance islands will grow wide enough that they must overlap each other in phase space, which is the situation shown in the rightmost plot of Figure \ref{fig::chirikov}. The Chirikov resonance overlap criterion says that this overlapping of resonance islands is exactly where regular circulating dynamics ends and large-scale chaotic transport across the phase space becomes possible. 

\begin{figure}
\includegraphics[width=0.4\columnwidth]{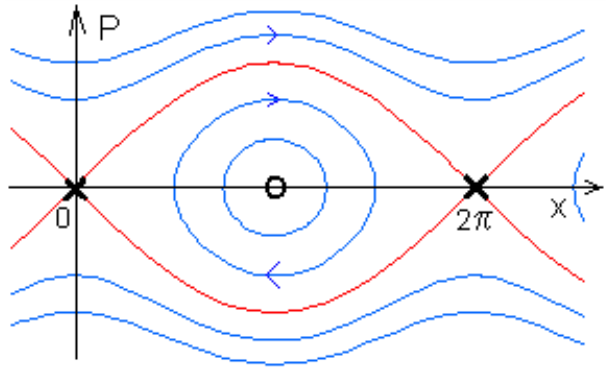}
\includegraphics[width=0.6\columnwidth]{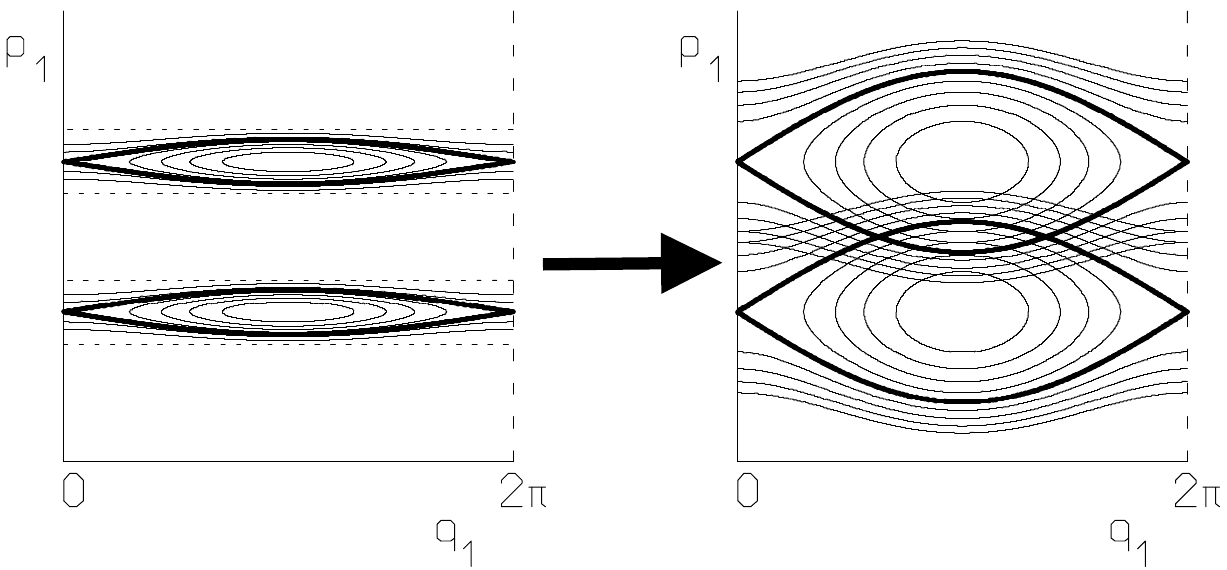}
\caption{ \label{fig::pendulum} Pendulum Phase portrait\cite{pendulumPhasePortrait} (left), growing and overlapping pendulums\cite{morbyBook} (right)}
\vspace{-18pt}
\end{figure}

To understand how this resonance overlapping and transition to global chaos occurs, it is instructive to discuss the dynamics inside each resonance island. In particular, it is known from Hamiltonian perturbation theory\cite{morbyBook} that around a resonance, if one uses a coordinate system which rotates with the resonance, the system phase portrait will resemble that of an undamped pendulum. The pendulum phase portrait is shown on the left of Figure \ref{fig::pendulum}; it contains three key types of motion: circulating invariant circles at the top and bottom of the plot, librating invariant circles in the middle (which do not go across the x-axis from 0 to $2\pi$), and separatrices (in red) between the circulating and librating regions. The separatrices are in fact stable and unstable manifolds of the unstable equilibrium point at $(0,0)=(2\pi,0)$ (since $x$ is an angle, these points are the same). There is also a stable equilibrium point $(\pi,0)$ at the center of the librating invariant circles.

If one does not use a coordinate frame rotating with the resonance, then for a 2D map like the standard map of Figure \ref{fig::chirikov} and a resonant rotation number $2\pi \frac{m}{n}$ with $m,n$ coprime, one essentially ends up with a pendulum-like phase portrait repeated $n$ times side by side inside that resonant island, except with each iteration of the map sending the $k$th pendulum onto the $(k+m)$th pendulum in the row (modulo $n$). Only after $n$ iterations will the map return each pendulum back onto itself. The stable and unstable pendulum equilibrium points of the previous paragraph will thus be replaced by stable and unstable periodic orbits, each having a period of $n$ map iterations, and the separatrices of the resonance will just become the stable and unstable manifolds of the unstable map-periodic orbit. Such a line of pendulums is visible, for example, in the island at $r/2\pi = 1/2$ in Figure \ref{fig::chirikov}, where two pendulum-shaped regions are clearly present. 

The above characterization of resonant islands indicates the dynamical mechanism by which their overlap generates global chaos and destroys regular motions. As illustrated on the right of Figure \ref{fig::pendulum}, when resonance islands overlap, their stable and unstable manifold separatrices must intersect, which facilitates transport from one resonance to the other. The reverse is also true; intersection of stable and unstable separatrices from different resonances implies resonance overlap. Furthermore, no circulating invariant circle can exist between two overlapping resonances in such a 2D system, since it would then have to cross at least one of these intersecting stable/unstable manifold separatrices, which is a contradiction\cite{olveraSimo}. 

\section{The internal structure of the 4:3 Orbit Family NHIM $\Xi_{\mu_{3}}$}

Though we used the standard map example to illustrate resonances and their overlapping, these phenomena also apply to other 2D symplectic, near-integrable maps on cylinders. Recall that just like the $K=0$ standard map, the $F_{\mu_{3}=0}$ (PCRTBP)-induced symplectic dynamics \emph{inside} the 2D cylindrical NHIM $\Xi_{\mu_{3}=0}$ are also integrable, since $\Xi_{\mu_{3}=0}$ is foliated entirely by $F_{\mu_{3}=0}$-invariant circles. Thus, as long as this NHIM, formed by the Jupiter-Ganymede 4:3 internal unstable resonant orbit family, persists into the CCR4BP, the previous discussion on resonances, overlapping, and destruction of circulating invariant circles applies the same to the dynamics inside the new perturbed $F_{\mu_{3}}$-invariant 2D NHIM $\Xi_{\mu_{3}}$ for $\mu_{3}>0$ as well. 

As described earlier, the NHIM $\Xi_{\mu_{3}=0}$ in the PCRTBP is foliated entirely by 4:3 Jupiter-Ganymede unstable PCRTBP flow-periodic orbits, which are also invariant circles of the PCRTBP stroboscopic map $F_{\mu_{3}=0}$. Given a point $\bold{x}_{0} \in \mathbb{R}^{4}$ belonging to a PCRTBP periodic orbit of period $T$, and letting $\phi(\bold{x}, t)$ be the time-$t$ flow of the point $\bold{x} \in \mathbb{R}^{4}$ by the PCRTBP equations of motion, we can define a function $K(\theta) = \phi(\bold{x}_{0}, T\frac{\theta}{2\pi})$. Since $K(0)=K(2\pi)$ due to the $T$-periodicity of the point $\bold{x}_{0}$, we can consider $\theta \in \mathbb{T}$. $K(\theta)$ is thus a parameterization of the $F_{\mu_{3}=0}$-invariant circle satisfying 
\begin{equation} \label{invarianceMu0} F_{\mu_{3}=0}(K(\theta)) = K(\theta+\omega)  \quad \quad \omega =  (2\pi/|\Omega_3-1|)( 2\pi/T)\end{equation} 
where $\omega$ is called the \emph{rotation number} of the invariant circle. 

Due to KAM theory\cite{delshams2008scattering}, the invariant circles inside $\Xi_{\mu_{3}=0}$ with $\frac{\omega}{2\pi}$ ``sufficiently irrational'' persist into the CCR4BP NHIM $\Xi_{\mu_{3}}$ for $\mu_{3}>0$ small enough. However, $F_{\mu_{3}=0}$-invariant circles with rational $\frac{\omega}{2\pi}$ disappear as soon as $\mu_{3}>0$, and are replaced by resonance islands (though an abuse of terminology, we henceforth refer to such $\omega$ as ``rational'' rotation numbers). As described in the previous section, each island will contain two $F_{\mu_{3}}$-periodic orbits as well as librational tori and separatrices. All these orbits will still be contained inside the 2D cylindrical NHIM $\Xi_{\mu_{3}}$, and  will thus inherit the NHIM's attached transverse stable and unstable directions. All objects inside each island are hence unstable when considering $F_{\mu_{3}}$ on its full phase space $\mathbb{R}^{4}$. However, if one considers the restriction of the dynamics of $F_{\mu_{3}}$ to only the 2D NHIM $\Xi_{\mu_{3}}$, then one of the two periodic orbits will be stable and the other unstable \emph{inside} the NHIM. We henceforth refer to these as \emph{NHIM-stable} and \emph{NHIM-unstable} periodic orbits. Separatrices will emanate from the latter.

\section{Computing Relevant Invariant Objects in the CCR4BP} \label{objectComputeSection}

In the previous sections, we discussed a variety of invariant objects which are important for understanding whether regular, circulating invariant circles (1D tori) can exist in a symplectic map on the 2D cylinder. These include the circulating invariant tori themselves, periodic orbits, and stable and unstable manifold separatrices emanating from those periodic orbits. Recall that we seek to investigate the dynamics induced by the CCR4BP stroboscopic map $F_{\mu_{3}}$ inside the 2D cylindrical NHIM $\Xi_{\mu_{3}}$ corresponding to the family of unstable 4:3 Jupiter-Ganymede resonant orbits. Thus, we must find these various invariant objects inside this NHIM (assuming the NHIM persists until the desired $\mu_{3}$, which we will show later to be the case). $\Xi_{\mu_{3}}$, however, is an \emph{unknown} 2D cylindrical submanifold of the 4D phase space of $F_{\mu_{3}}$. Thus, even if one can theoretically interpret $F_{\mu_{3}}$ to be a map restricted to a 2D cylinder, due to the lack of previous knowledge on $\Xi_{\mu_{3}}$'s location, in practice all the invariant circles, map-periodic orbits, and separatrices inside $\Xi_{\mu_{3}}$ must be computed not in a 2D cylindrical phase space but in the full $\mathbb{R}^{4}$ phase space of the stroboscopic map. 

To compute the invariant circles, map-periodic orbits, and stable and unstable separatrices inside the unstable 4:3 Jupiter-Ganymede orbit family NHIM for the CCR4BP stroboscopic map, various computational methods are needed. Some new, improved, and efficient methods for computing invariant circles (1D tori) in such cases, as well as their stable and unstable manifolds, were developed in our previous work\cite{kumar2022}. And over the course of this study, we were also able to extend the aforementioned invariant circle computation methods to the case of map-periodic orbits, and the torus stable/unstable manifold computation algorithm to the case of separatrices inside a NHIM. We summarize these three methods in this section, though a full explanation of our new periodic orbit and separatrix computation methods will require a separate paper\cite{kumar2023upcoming} (in preparation). Readers more interested in dynamical results than methodology may skip this section without issue. 

\subsection{Parameterization Methods for Computing Invariant Circles (Tori)} \label{quasiNewton}

Though Equation \eqref{invarianceMu0} was for $\mu_{3}=0$, it is in fact also the equation which must be solved to compute those invariant circles $K$ which persist for $\mu_{3} >0$. In this $\mu_{3}>0$ case, it takes the form
\begin{equation} \label{invariance} F_{\mu_{3}}(K(\theta)) = K(\theta+\omega)  \quad \quad \omega \text{ inherited from the corresponding } \mu_{3}=0 \text{ torus} \end{equation} 
In previous work\cite{kumar2022}, we developed a quasi-Newton method for solving Eq. \eqref{invariance} for unstable tori in periodically-forced PCRTBP models such as the CCR4BP. The method solves not only for $K(\theta)$, but also simultaneously finds a matrix-valued function $P(\theta): \mathbb{T} \rightarrow \mathbb{R}^{4 \times 4}$ and $\Lambda \in \mathbb{R}^{4 \times 4}$ satisfying
		 \begin{equation}  \label{bundleEquations} DF_{\mu_{3}}(K(\theta)) P(\theta) = P(\theta+\omega) \Lambda \end{equation} 
where we mandate $\Lambda$ to be the unstable torus' Floquet stability matrix, of form
\begin{equation}  \label{Lambdaform}
\Lambda=\begin{bmatrix}
1 &  T & 0 & 0 \\ 0 &  1   & 0 & 0 \\ 0 & 0  & \lambda_s  & 0 \\ 0 &  0 & 0 & \lambda_u \end{bmatrix}
 \end{equation}
For each $\theta \in \mathbb{T}$, the four columns of $P(\theta)$ are the linearly independent tangent, center, stable, and unstable directions of the torus at the point $K(\theta)$, in that order. $T \in \mathbb{R}$ is a shear constant and $\lambda_s, \lambda_u \in \mathbb{R}$ are the constant stable/unstable multipliers for the circle ($\lambda_s<1$, $\lambda_u > 1$). The torus tangent and center directions will both be tangent to the 2D NHIM $\Xi_{\mu_{3}}$ containing the torus. 

As it turns out, solving simultaneously for $K$, $P$, and $\Lambda$  not only gives stability information, but actually has lower computational complexity than more commonly-used methods which solve for $K$ alone \cite{kumar2022}. The method converges given a sufficiently accurate initial guess, so one can use it to numerically continue invariant circles from the PCRTBP (where they are easily computed) to the $\mu_{3}>0$ CCR4BP. Once a solution $K,P,\Lambda$ is computed at some $\mu_{3}$, it is possible to also compute $\frac{dK}{d\mu_{3}}$ using a Lindstedt method to improve the initial guess for the next continuation step, as described in Kumar et al \cite{kumar2021aug}. During the continuation, due to the changing $\mu_{3}$ and Kepler's third law, one must vary $r_{13}$ in order to keep $\Omega_p$ and $\omega$ constant (which is needed). 



Finally, we note that CCR4BP stroboscopic map $F_{\mu_{3}}$-invariant circles correspond to 2D tori of the CCR4BP flow. Any such 2D torus can be parameterized in the CCR4BP flow's 5D phase space as a function of 2 angles, $K_2(\theta,\theta_{p}) : \mathbb{T}^{2} \rightarrow \mathbb{R}^{4} \times \mathbb{T}$, where $\theta_{p}$ is the perturbation phase angle from the CCR4BP flow phase space. Recall that the map $F_{\mu_{3}}$ was defined by fixing $\theta_{p}=\theta_{p,f}$; thus, the $F_{\mu_{3}}$-invariant circle $K(\theta)$ is related to its flow-invariant counterpart through the equation $K(\theta) = K_{2}(\theta, \theta_{p,f})$. At times, however, it can be beneficial to fix $\theta=\theta_{f}$ and then compute $K_{P}(\theta_{p})=K_{2}(\theta_{f}, \theta_{p}):\mathbb{T} \rightarrow  \mathbb{R}^{4} \times \mathbb{T}$ instead. In Kumar et al\cite{kumar2023}, we extended the quasi-Newton method for computing stroboscopic map-invariant $K$ to the computation of $K_{P}$ as well. 

\subsection{Computing Long Periodic Orbits} \label{poCompute}

As explained earlier, separatrices attached to the NHIM-unstable $F_{\mu_{3}}$-periodic orbits drive resonance overlapping and destruction of invariant circles. Computing separatrices, however, first requires finding the NHIM-unstable periodic orbits from which they emanate. For a resonance at rational $\omega=2\pi \frac{m}{n}$, $m,n \in \mathbb{Z}$ coprime, the period of its periodic orbit will be $n$ iterations of $F_{\mu_{3}}$. But if $n$ is large, as is the case for the orbits we study later, the resulting $n$-periodic orbit will be very difficult to compute using single-shooting. Thus, a multiple shooting algorithm is required. For an $n$-periodic orbit, we will find $n$ multiple shooting points $X(k) \in \mathbb{R}^{4}$, $k=0,1,\dots, n-1$, satisfying
\begin{equation} \label{poSolve} F_{\mu_{3}}(X(k)) = X(k+m \mod n) \end{equation}
This is equivalent to the slightly different equation $F_{\mu_{3}}(X(k)) = X(k+1 \mod n)$ often seen in the literature\cite{parkerAnderson}; only the ordering of the solution points $X(k)$ will be different. We choose the ordering of $X(k)$ induced by Equation \eqref{poSolve} because for the resonant island at $\frac{\omega}{2\pi}=\frac{m}{n}$, points of its solution $X(k)$ at consecutive $k$ will then also lie physically next to each other inside the NHIM $\Xi_{\mu_{3}}$. 

Multiple-shooting algorithms to solve Equation \eqref{poSolve} have been developed and used previously\cite{parkerAnderson} in many astrodynamics applications; those algorithms however involve using Newton's method to solve for a vector containing all of the $X(k)$ components for all $k$-values, which requires solving a large-dimensional linear system of equations in each Newton step. However, in this study, we developed a new multiple-shooting quasi-Newton method that avoids solving large linear systems. To do this, we augment Equation \ref{poSolve} with another equation to be solved simultaneously for $P(k): \{0,1,\dots,n-1\} \rightarrow \mathbb{R}^{4 \times 4}$ and $\Lambda \in \mathbb{R}^{4 \times 4}$ satisfying
\begin{equation}  \label{poBundleEquations} DF_{\mu_{3}}(X(k)) P(k) = P(k+m \mod n) \Lambda \end{equation}

For $\mu_{3}=0$, $\Lambda$ is mandated to have the form given in Equation \eqref{Lambdaform}, while the columns of $P$ will be as described in the previous section on tori. However, for $\mu_{3}>0$, $\Lambda$ is required to instead be diagonal with entries $\lambda_{1}$, $\lambda_{2}$, $\lambda_{s}$, and $\lambda_{u}$; if $X(k)$ is a NHIM-unstable orbit, the third and fourth columns of $P$ will continue to be its stable and unstable directions \emph{transverse} to the NHIM, but the first and second columns will be the new stable and unstable directions \emph{tangent} to the NHIM which are generated by the appearance of the resonant island. In this case, $\lambda_{1}<1$ and $\lambda_{2}>1$ will both be close to 1, while $\lambda_{s}<1$ and $\lambda_{u}>1$ will be further from unity. As is explained in our torus quasi-Newton method paper\cite{kumar2022}, that method uses $P$ and $\Lambda$ to make each torus quasi-Newton step upper triangular (in fact, almost diagonal); the exact same is also possible using $P$ and $\Lambda$ in the multiple-shooting periodic orbit quasi-Newton method. 

Though the multiple-shooting algorithm is inspired by our torus computation method, some modifications to the torus algorithm were required to make it work for periodic orbits. Though we refer the reader to our upcoming paper\cite{kumar2023upcoming} for the full details, we will briefly mention one important change, which is in the solution of cohomological equations. For the torus case, these equations were of form $\lambda_{a} a(\theta) - \lambda_{b} a(\theta+\omega) = b(\theta)$, with $\lambda_{a}, \lambda_{b}$, and $b:\mathbb{T} \rightarrow \mathbb{R}$ known; each torus quasi-Newton step requires solving 16 such equations, which can be done using Fourier series or fixed-point iteration\cite{kumar2022}. However, in the periodic orbit case, the equations become $\lambda_{a} a(k) - \lambda_{b} a(k+m \mod n) = b(k)$. If $\frac{\lambda_{a}}{\lambda_{b}}$ is not extremely close to 1, then the same fixed point iteration from the torus case\cite{kumar2022} still works. Otherwise, if $\lambda_{a}$ and $\lambda_{b}$ are nearly but not exactly equal, then one can instead derive an explicit formula for $a(k)$ by taking advantage of the periodicity of the index $k$. If $\lambda_{a}=\lambda_{b}$, then given any solution $a(k)$, $a(k)+C$ is also a solution for any $C \in \mathbb{R}$; thus, we can arbitrarily set $a(0)=0$ and use the relation $a(k+m \mod n) = a(k) - b(k)/\lambda_{a} $ to determine all the other $a(k)$. 

\subsection{Computing Separatrices} \label{separatrixCompute}

Once a solution $X,P,\Lambda$ to Equations \eqref{poSolve}-\eqref{poBundleEquations} has been found at a resonance $\omega=2\pi\frac{m}{n}$ with $m,n \in \mathbb{Z}$ coprime, then if $X$ is a NHIM-unstable periodic orbit, we would like to compute the separatrices emanating from it. These 1D separatrices will be comprised of the stable and unstable manifolds of this periodic orbit \emph{inside} the 2D NHIM $\Xi_{\mu_{3}}$. The information given by $P(k)$ and $\Lambda$ on stable/unstable directions and multipliers can be used to compute these separatrices; the first and second columns of $P(k)$ give the linear approximations to the stable and unstable separatrices emanating from each point $X(k)$, while the first two diagonal entries $\lambda_{1}<1$ and $\lambda_{2} >1$ of $\Lambda$ give their corresponding (weak, near unity) stable and unstable multipliers. 

Since these separatrices correspond to weak expansion and contraction inside the NHIM $\Xi_{\mu_{3}}$ (as opposed to the strong expansion/contraction transverse to the NHIM), globalizing the aforementioned linear stable/unstable separatrix approximations using numerical integration will lead to large numerical errors; the separatrices we want are contained inside the NHIM $\Xi_{\mu_{3}}$, but integration will amplify even very small errors in the linear approximation in the direction of the NHIM's strong, transverse unstable direction, pushing the computed separatrix off of the NHIM. Thus, we instead will use a parameterization method\cite{haroetal} to compute the separatrices as polynomial curves, including nonlinear terms. For this, we find a function $W(k, s): \{0,1,\dots,n-1\} \times \mathbb{R} \rightarrow \mathbb{R}^{4}$ satisfying 
\begin{equation}  \label{invariancequationWfinal}   F_{\mu_{3}}(W(k, s)) = W(k + m \mod n, \lambda s) \end{equation}
where $\lambda=\lambda_{1}$ or $\lambda_{2}$ depending on which separatrix is sought. Equation \eqref{invariancequationWfinal} can be solved recursively by expressing $W$ as a set of $n$ Taylor series depending on the integer $k$
\begin{equation}  \label{taylorSeries}   W(k, s) = \sum_{i=0}^{\infty} W_{i}(k) s^{i} \end{equation}
where $W_{0}(k)$ is the known function $X(k)$ and $W_{1}(k)$ is known from column 1 or 2 of $P(k)$. 

The method of computing each $W_{i}$ is very similar to the corresponding method for stable and unstable manifolds of tori given in our previous paper\cite{kumar2022}. Each term $W_{i}(k)$ for $i\geq2$ can be found, given all the terms $W_{j}(k)$ with $j<i$, by letting $W_{i}(k)=P(k)V_{i}(k)$ and solving the equation
\begin{equation}  \label{WiTermsolve}   \Lambda V_{i}(k) - \lambda^{i} V_{i}(k + m \mod n) = -P(k)^{-1}E_{i}(k)\end{equation}
where $E_{i}(k)$ is the $s^{i}$ Taylor coefficient of $E(k,s)=F_{\mu_{3}}\left(\sum_{j=0}^{i-1} W_{j}(k) s^{j}\right) - \sum_{j=0}^{i-1} W_{j}(k+m \mod n) (\lambda s)^{j}$; $E(k,s)$ is just the error in Equation \eqref{invariancequationWfinal} when the degree $i-1$ truncation of $W$ is substituted in its place. The coefficient $E_{i}$ can be computed using jet transport\cite{perezpalau2015}, exactly like in the torus case\cite{kumar2022}. For full details of this method, we again refer readers to our upcoming paper\cite{kumar2023upcoming}. In practice, these series will be truncated at some finite degree in $s$. Also, they are valid only on a finite interval of $s$ values centered at 0, but we will find this to be enough for our study; this interval can be found by finding the maximum value of $|s|$ for each $k$ such that Equation \eqref{invariancequationWfinal} is true within some fixed tolerance. In this study, we use a tolerance of $10^{-5}$. 

\section{Evidence of Secondary Resonance Overlap inside the 4:3 J-G Family} 

Given the previous information on the NHIM $\Xi_{3}$ formed by the Jupiter-Ganymede (J-G) unstable 4:3 resonant orbits, on resonance islands inside $\Xi_{3}$, and on resonance overlapping, we can now discuss the properties of the 4:3 J-G family in the CCR4BP. In a previous paper\cite{kumar2023} on potential transfers from J-G to Jupiter-Europa resonances, we attempted to compute a variety of unstable J-G 4:3 orbits in the CCR4BP. In that work, the only method used for this was continuation of orbits by $\mu_{3}$ from the J-G PCRTBP as invariant circles (at irrational $\frac{\omega}{2\pi}$) into the Jupiter-Europa-Ganymede (J-E-G) CCR4BP; however, continuation implicitly assumes that those circles persist until Europa's physical $\mu_{3}=\mu_{3,e} \approx 2.52651 \times 10^{-5}$, which as we will see may not be the case. Using that methodology, only low-energy J-G 4:3 orbits were successfully continued as tori until the J-E-G CCR4BP $\mu_{3}=\mu_{3,e}$. Attempts to continue the higher-energy PCRTBP unstable 4:3 orbits, whose manifolds are more unstable and thus useful for low-TOF transfers, into the CCR4BP stopped before $\mu_{3,e}$. 

\begin{figure}
\begin{centering}
\includegraphics[width=0.49\columnwidth]{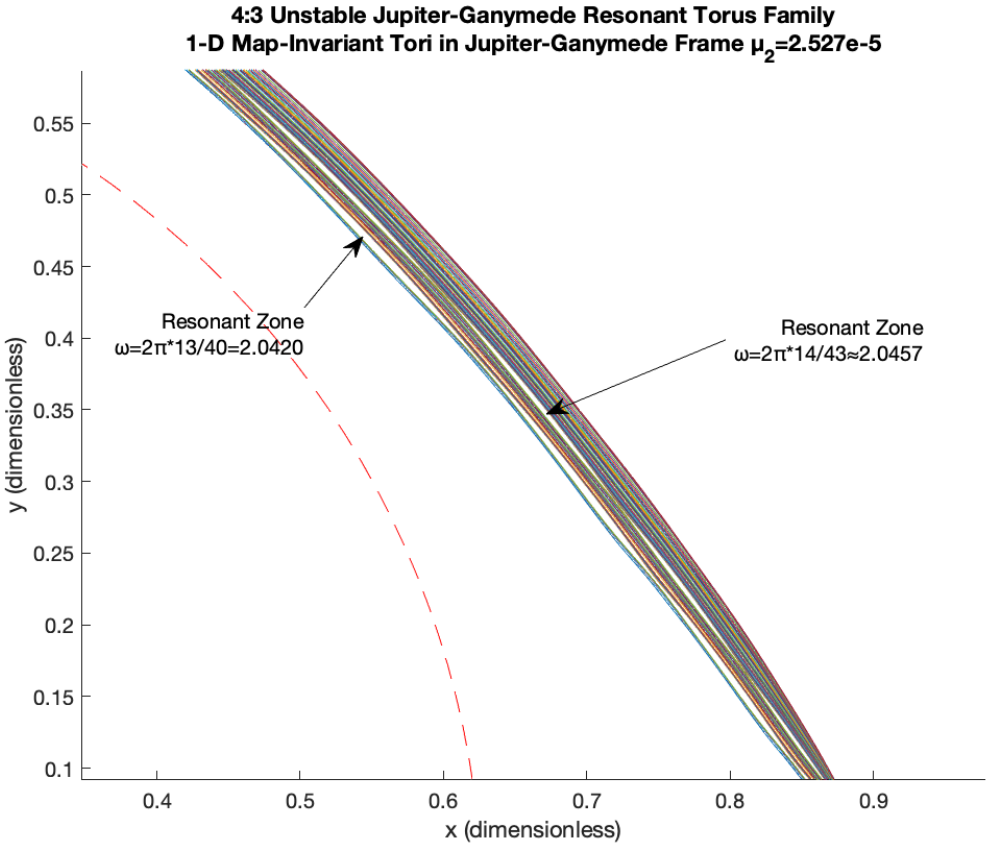}
\includegraphics[width=0.5\columnwidth]{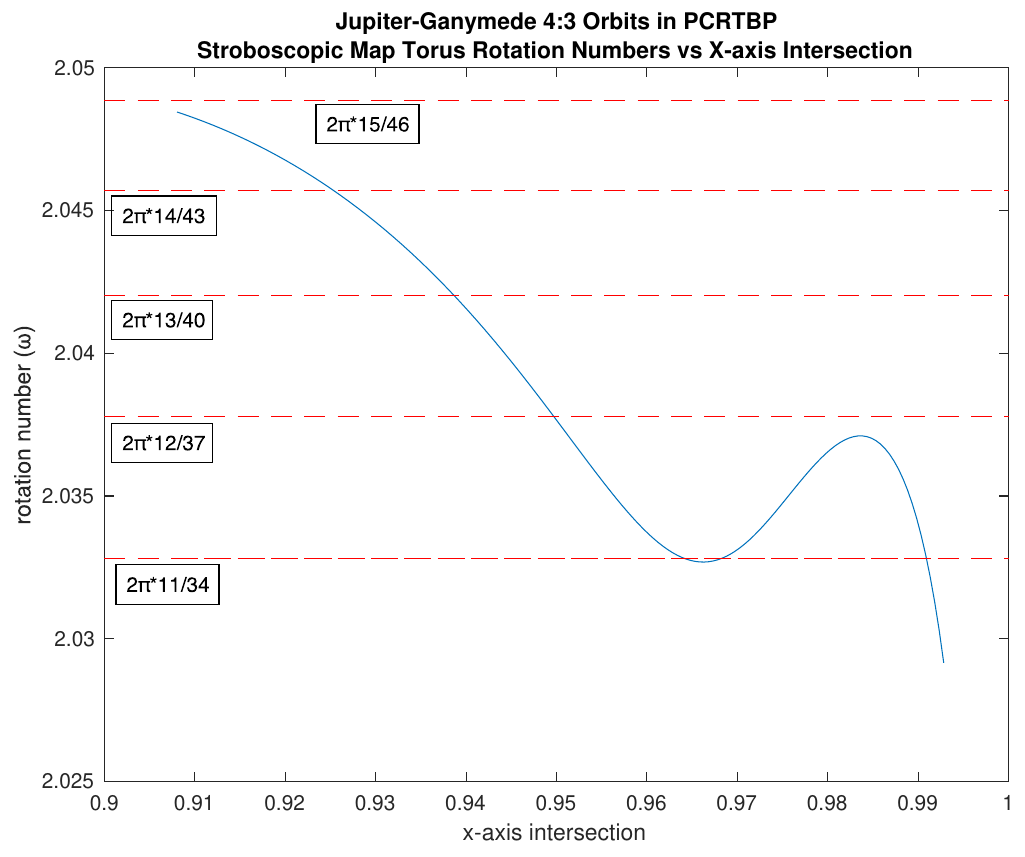}
\caption{ \label{fig:43resonances} Resonant Gaps Within 4:3 Tori at $\mu_{3}=\mu_{3,e}$ (left\cite{kumar2023}), 4:3 J-G Family $\omega$ values (right\cite{kumar2023})}
\vspace{-18pt}
\end{centering}
\end{figure}

The lower-energy tori which were successfully continued to the physical system nevertheless showed an interesting behavior. In particular, between the successfully computed tori, two significant gaps appeared, both shaped like strings of pendulum phase portraits similar to the left of Figure \ref{fig::pendulum}. These tori, with clearly visible gaps, are shown on the left of Figure \ref{fig:43resonances}. Furthermore, these gaps occur at $\omega$ values of $2\pi \frac{13}{40}$ and $2\pi \frac{14}{43}$, making it obvious that these gaps are in fact resonant islands inside the unstable 4:3 J-G family NHIM $\Xi_{\mu_{3}}$. Finding all possible $\omega$ values for the unstable 4:3 J-G family, determined using the $\mu_{3}=0$ PCRTBP orbit family and Equation \eqref{invarianceMu0}, we found that $\omega$ ranged from about 2.0485 to 2.028. In this interval, there are four $\frac{\omega}{2\pi}$ values with denominator less than 50: $\frac{\omega}{2\pi} = \frac{14}{43}, \frac{13}{40}, \frac{12}{37}$, and $\frac{11}{34}$. The right of Figure \ref{fig:43resonances} plots $\omega$ versus the vertical positive $x$-axis crossing coordinate of each PCRTBP 4:3 J-G orbit, with those rational $\omega$ overlaid. 

Given the significant resonant islands observed among even the successfully computed tori, the presence of lower-denominator rational $\omega$ values in $\omega \in [2.028, 2.0405]$ (whose corresponding tori all failed to continue until $\mu_{3}=\mu_{3,e}$), and the fact that the failed continuation orbits all make closer passes of Europa's orbit than those whose continuations succeeded (meaning a stronger perturbative effect from $\mu_{3}$), a reasonable hypothesis is that the continuation failure could be due to resonant islands in the range $\omega \in [2.028, 2.0405]$ overlapping. As discussed earlier, overlapping resonances in a 2D map (such as the internal dynamics of the 4:3 J-G family NHIM $\Xi_{\mu_{3}}$) necessarily destroy all invariant circles between them. To investigate whether this is the case, we first use a torus continuation based method to study the dynamics inside $\Xi_{\mu_{3}}$. 

Before describing the methodology and results of this investigation, we would like to make a note about terminology. In the previous discussion, we referred to resonant islands inside the unstable 4:3 J-G family NHIM $\Xi_{\mu_{3}}$ at rational $\omega$ values. However, note that the 4:3 J-G orbits which comprise $\Xi_{\mu_{3}}$ are themselves \emph{all} in a mean motion resonance (MMR) with Ganymede, regardless of $\omega$, and are in fact all inside a higher-dimensional resonance island of the wider phase space $\mathbb{R}^{4}$. Thus, to distinguish from the 4:3 J-G MMR, the rational rotation numbers $\omega$ and their corresponding resonant islands inside $\Xi_{\mu_{3}}$ should be referred to as \emph{secondary} resonances. To understand the physical meaning of a secondary resonance, one can rearrange the expression given for $\omega$ in Equation \eqref{invarianceMu0}; namely, one finds that $\frac{\omega}{2\pi} =  \frac{2\pi/|\Omega_3-1|}{T}$. But $2\pi/|\Omega_3-1|$ is just the synodic period of Europa's revolution with respect to the J-G frame, while $T$ is the period of some 4:3 J-G unstable resonant periodic orbit from the J-G PCRTBP. Thus, at a secondary resonance where $\frac{\omega}{2\pi}$ is rational, the period of that J-G periodic orbit is a rational multiple of the J-G-synodic period of Europa's revolution.  

\subsection{Torus Continuation-Based Secondary Resonance Investigation: Methodology} 

In reality, $\mu_{3}$ has a fixed value $\mu_{3,e}$ for Europa, which is thus the only value of $\mu_{3}$ of practical interest for real missions. However, studying the behavior of the orbits inside $\Xi_{\mu_{3}}$ as a function of $\mu_{3}$ ranging from $0$ to $\mu_{3,e}$ can help us predict what the orbit family properties will be at $\mu_{3}=\mu_{3,e}$. Our torus continuation-based investigation operates in this spirit, taking inspiration from Figure \ref{fig:43resonances}, where secondary resonant islands inside $\Xi_{\mu_{3}}$ for $\mu_{3}=\mu_{3,e}$ became visible through the continuation of the invariant circles surrounding them. Although no invariant circles with $\omega \in [2.028, 2.0405]$ persisted until $\mu_{3}=\mu_{3,e}$, we can instead continue circles to smaller $\mu_{3}$ values, and see if secondary resonant islands appear among the tori which persist at those $\mu_{3}$ values. Doing the same for a sequence of increasing $\mu_{3}$ values should provide information about how the size of the islands grows with $\mu_{3}$, which is key to determining if they overlap. 

We carry out this procedure starting from a fine grid of higher-energy, 4:3 J-G PCRTBP orbits from the right plot of Figure \ref{fig:43resonances} lying between $x \approx 0.942, \omega \approx 2.0405$ and the plot local minimum at $x \approx 0.965, \omega \approx 2.032685$; all these orbits failed to continue until $\mu_{3,e}$. Using our quasi-Newton method described earlier, we try to continue each $\mu_{3}=0$ invariant circle from the aforementioned grid to $\mu_{3}$ values of $\mu_{3,j}=10^{-7}j$ for $j=1, 2, \dots, 253$, going as high in $\mu_{3}$ as possible without dropping below a minimum continuation step size. We start with a continuation step size of $10^{-7}$, followed by $10^{-8}$ and finally $10^{-9}$ as the minimum step size. 

The fine grid of 4:3 J-G PCRTBP orbits from which we start the continuation corresponds to a set of rotation numbers $\omega_{k,S} \in [2.032685, 2.0405]$. However, tori at certain $\omega$ values will persist longer than others; in order to avoid misleadingly wide-looking secondary resonant islands between persisting tori, the initial grid of orbits should be chosen so that they persist until values of $\mu_{3}$ as large as possible. Prior research\cite{macKay1992} suggests that tori with noble $\omega$, that is, a continued fraction expansion\cite{weissteinCF} of form $[a_{0}, a_{1}, \dots, a_{n}, 1, 1, 1, \dots]$, $a_{i} \in \mathbb{Z}^{+}$ for all $i$, survive longest. To get a fine grid of noble rotation numbers $\omega_{k, S} \in [2.032685, 2.0405]$, start with an evenly spaced grid of $\omega_{k} \in [2.032685, 2.0405]$, and then: \vspace{-5pt}
	    \begin{enumerate}
	    \setlength\itemsep{-0\parsep}
	    \item Compute the (finite, since on a computer) continued fraction (CF) expansion of $\omega_{k}$ 
	    \item Starting with $j=1$, replace the last $j$ CF terms with 1. Convert this CF to a number $\omega_{k,j}$, check if $|\omega_{k,j} - \omega_{k}|>$ tol (we use $\frac{\omega_{k+1}-\omega_{k}}{2}$), and if not, increase $j$ by 1 and repeat this step
	    \item Once $|\omega_{k,j} - \omega_{k}|>$ tol, repeatedly increment the $j$th from last CF term by 1 and recompute the resulting $\omega_{k,j}$ until $|\omega_{k,j} - \omega_{k}|<$ tol again. The resulting $\omega_{k,j}$ is $\omega_{k, S}$. 
	    \end{enumerate} \vspace{-5pt}
Doing this for all $\omega_{k}$, we get the grid of noble $\omega_{k, S}$ values corresponding to the tori from which we should start the continuation. We take PCRTBP orbits at these $\omega_{k, S}$ values and continue them. 

\begin{figure}
\begin{centering}
\includegraphics[width=\columnwidth]{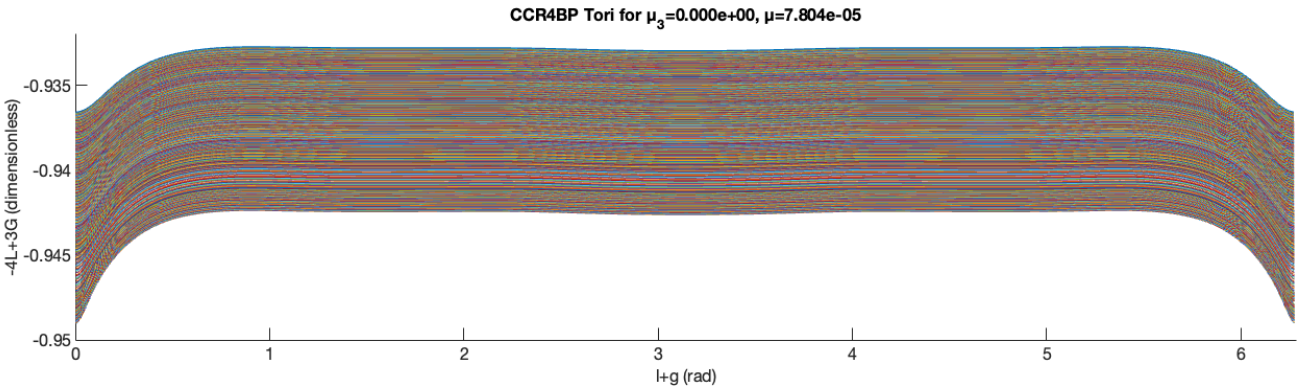}
\caption{4:3 J-G PCRTBP orbit family for $\omega \in [2.032685, 2.0405]$ plotted in $P_{2}$ vs $-Q_{2}$ coordinates}
\label{fig::pcrtbpTori} \vspace{-18pt}
\end{centering}
\end{figure}

To visualize the results, recall from the section on NHIMs that a 2D plot of the persisting tori at each $\mu_{3,j}$ value is enough, since all persisting tori will belong to the 2D cylindrical NHIM $\Xi_{\mu_{3,j}}$. Nevertheless, one can choose a particularly advantageous set of 2D coordinates for plotting by considering the fact these are unstable 4:3 J-G mean motion resonant orbits. In particular, though the tori are all computed in Cartesian coordinates, we can first transform the resulting points to Jupiter-centric, J-G synodic Delaunay coordinates\cite{celletti} $(L,G,\ell,g)$. Here, $\ell$ is the osculating spacecraft mean anomaly, while $g$ is its osculating \emph{J-G synodic} argument of periapse. As $g$ is measured with respect to the synodic J-G frame $x$-axis, $\dot g$ will be approximately -1 times Ganymede's mean motion around Jupiter, even though (in fact, since) the spacecraft argument of periapse remains near-constant in an inertial reference frame. All this becomes exact in the $\mu=\mu_{3}=0$ case (rotating Kepler problem). Thus, carrying out the linear canonical transformation
\begin{gather} Q_{1}=3\ell+4g \quad \quad Q_{2}=-\ell-g  \quad \quad P_{1}= -L+G \quad  \quad P_{2} =-4L+3G \end{gather}
we have that $Q_{1}$ is a ``slow angle'' for the 4:3 J-G MMR, since by definition $\dot Q_{1}=0$ inside the 4:3 J-G resonance in the $\mu=\mu_{3}=0$ Keplerian case. $Q_{2}$ on the other hand is a ``fast angle'' inside the 4:3 J-G MMR, with $\dot Q_{2}<0$ even for $\mu=\mu_{3}=0$. As it turns out, plotting all the torus points in the 2D space $(Q_{2}, P_{2})$ leads to a very simple and useful representation, where in the PCRTBP case ($\mu_{3}=0$), all unstable 4:3 J-G orbits are near-horizontal curves showing clearly circulating invariant circles. This is shown in Figure \ref{fig::pcrtbpTori} for the 4:3 J-G PCRTBP  orbits with $\omega \in [2.032685, 2.0405]$. 

We briefly explain why these coordinates lead to simple plots. From the perturbation theory of Hamiltonian systems\cite{morbyBook}, in the PCRTBP (since a flow is required for the following), having $\dot Q_{2}>0$ near the 4:3 MMR even when $\mu=0$ means that we can find a near-identity canonical transformation $(Q_{1}, Q_{2}, P_{1}, P_{2}) = \chi(\bar Q_{1}, \bar Q_{2}, \bar P_{1}, \bar P_{2})$, valid around the 4:3 MMR, which essentially averages the Hamiltonian over the fast motion of $Q_{2}$. The resulting PCRTBP Hamiltonian is independent of $\bar Q_{2}$, which  makes $\bar P_{2}$ an integral of motion and thus decouples the motion of $(\bar Q_{2}, \bar P_{2})$ from $(\bar Q_{1}, \bar P_{1})$, which then follow the pendulum phase portrait of Figure \ref{fig::pendulum}. In the averaged coordinates, our unstable 4:3 J-G orbits occur at the unstable equilibrium point in $(\bar Q_{1}, \bar P_{1})$ space, which is not interesting. But in averaged $(\bar Q_{2}, \bar P_{2})$ coordinates, each 4:3 orbit will just be a horizontal circulating invariant torus rotating in the $\bar Q_{2}$ direction, due to the integral of motion $\bar P_{2}$. Since $\chi$ was a near-identity (though complicated) transformation, the orbit plots in non-averaged $(Q_{2}, P_{2})$ are similar. 

\subsection{Torus Continuation-Based Secondary Resonance Investigation: Results} 

\begin{figure}
\begin{centering}
\includegraphics[width=\columnwidth]{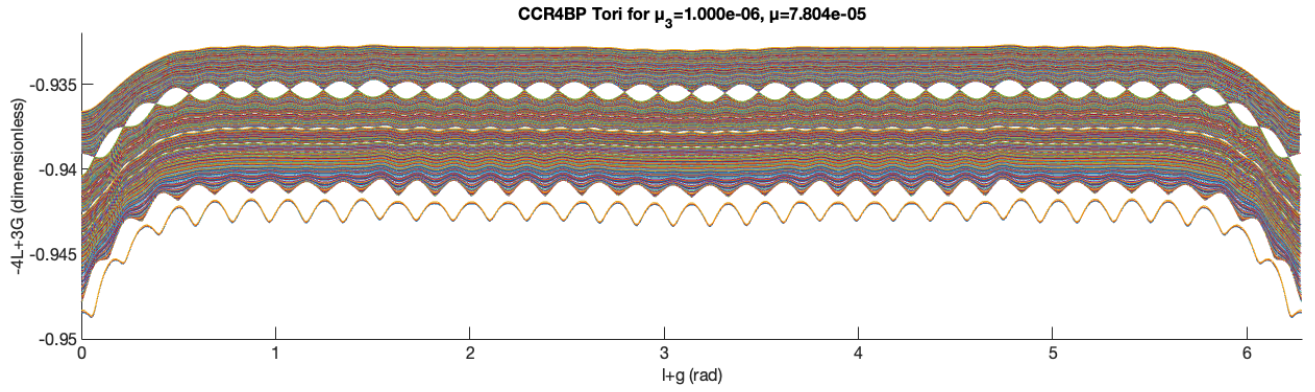}
\includegraphics[width=\columnwidth]{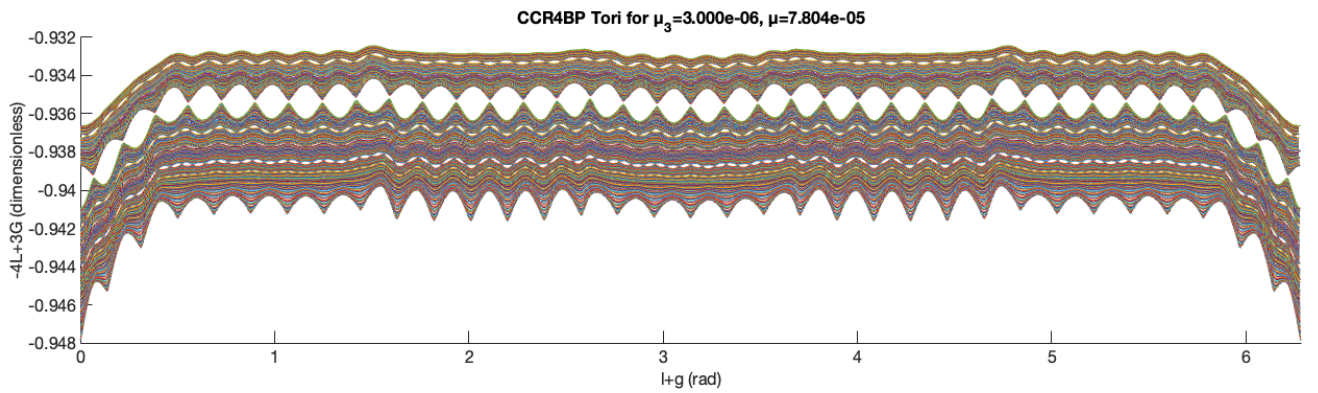}
\includegraphics[width=\columnwidth]{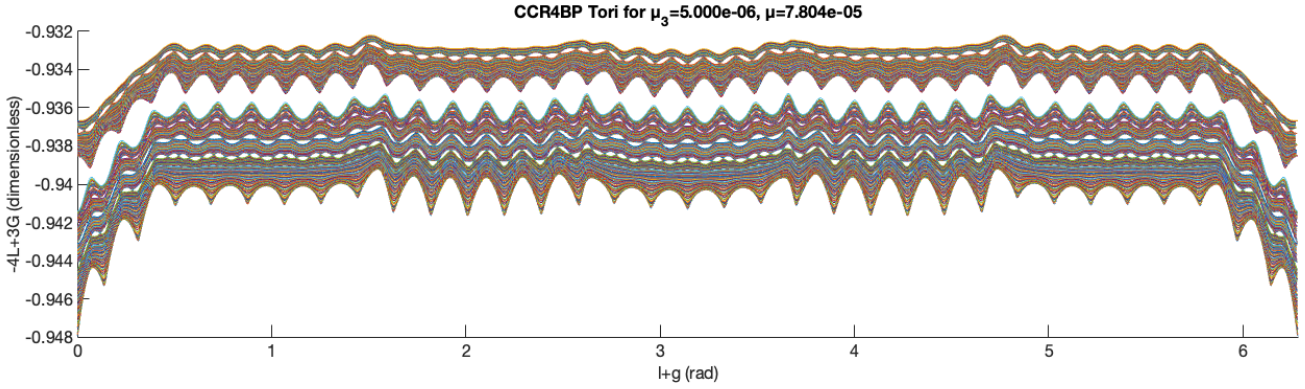}
\includegraphics[width=\columnwidth]{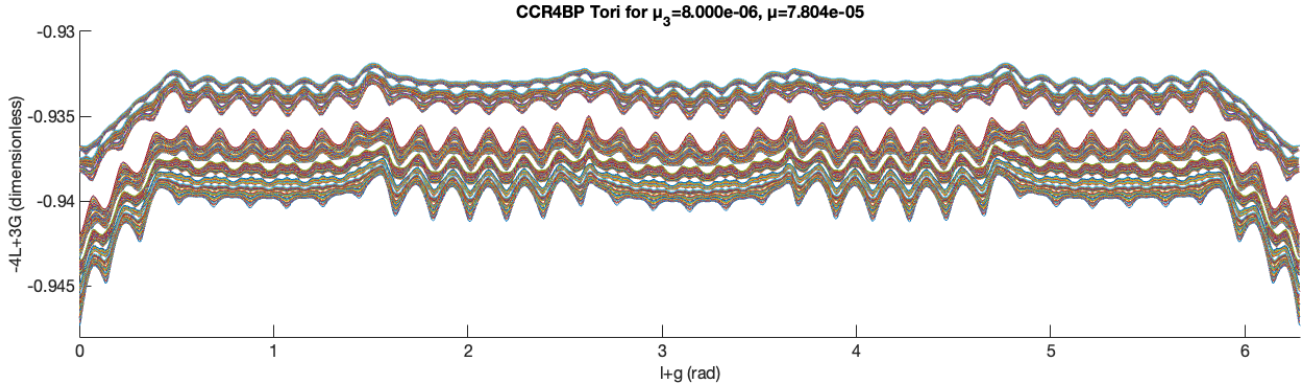}
\includegraphics[width=\columnwidth]{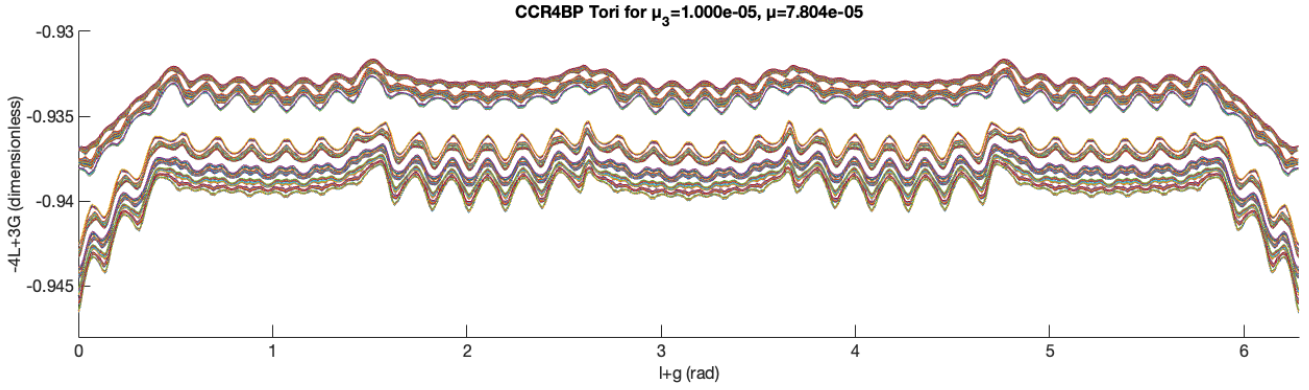}
\caption{Persisting CCR4BP 4:3 J-G orbit tori for $\omega \in [2.032685, 2.0405]$ plotted in $(-Q_{2}, P_{2})$ at $\mu_{3} \times 10^{6}=1, 3, 5, 8, 10$ (from top to bottom) }
\label{fig::continuations1} \vspace{-18pt}
\end{centering}
\end{figure}

\begin{figure}
\begin{centering}
\includegraphics[width=\columnwidth]{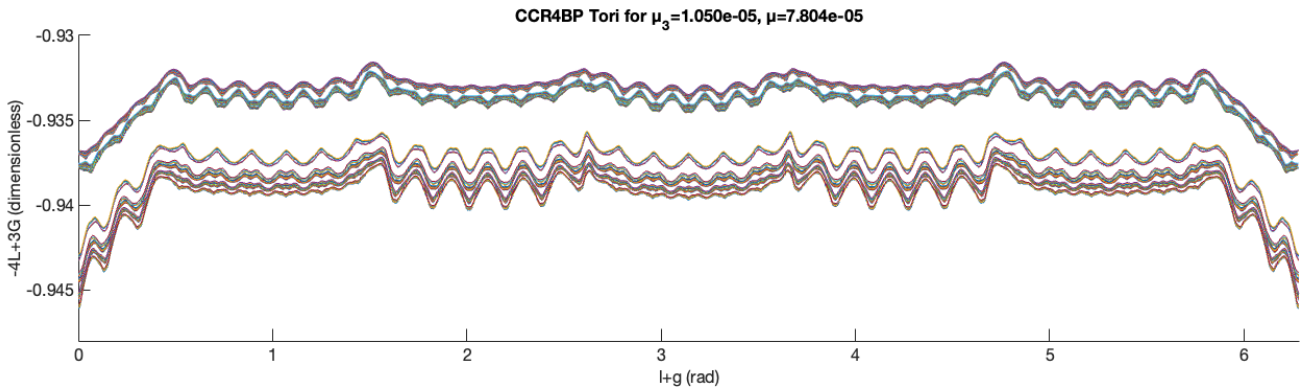}
\includegraphics[width=\columnwidth]{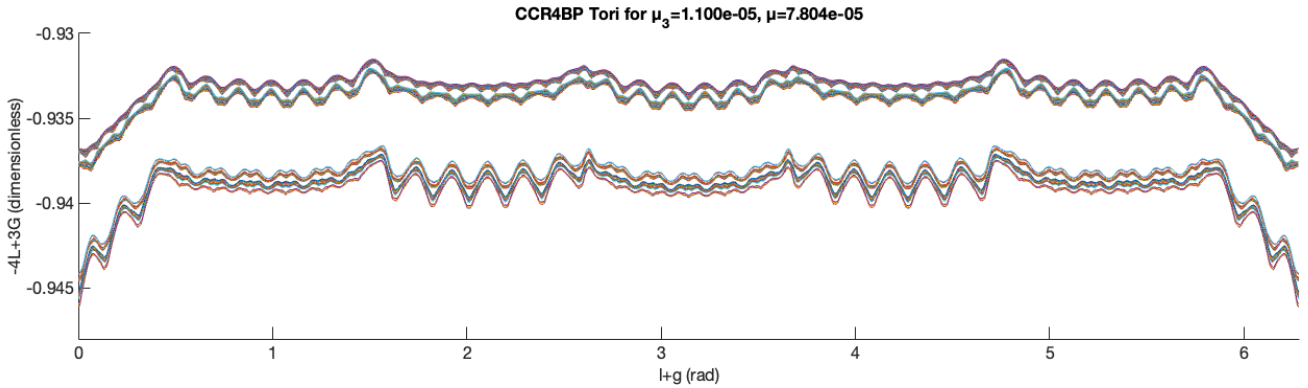}
\includegraphics[width=\columnwidth]{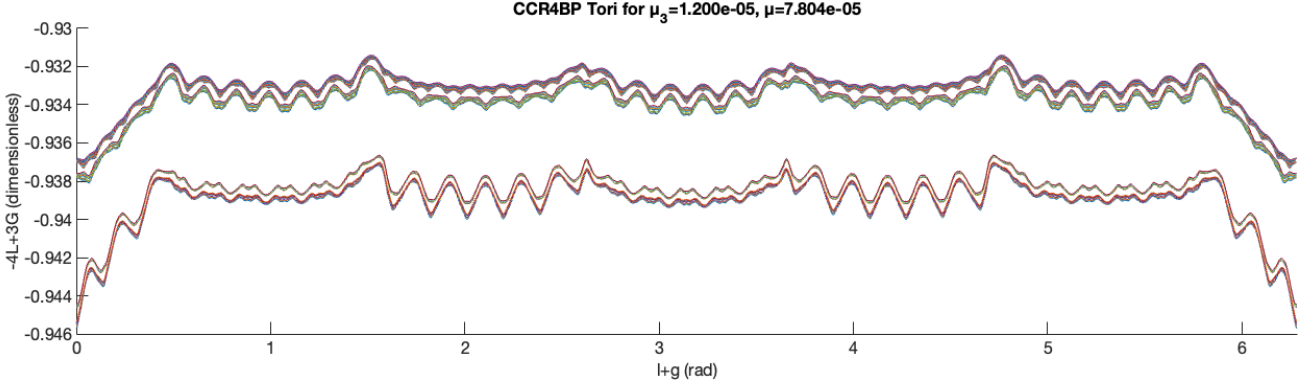}
\includegraphics[width=\columnwidth]{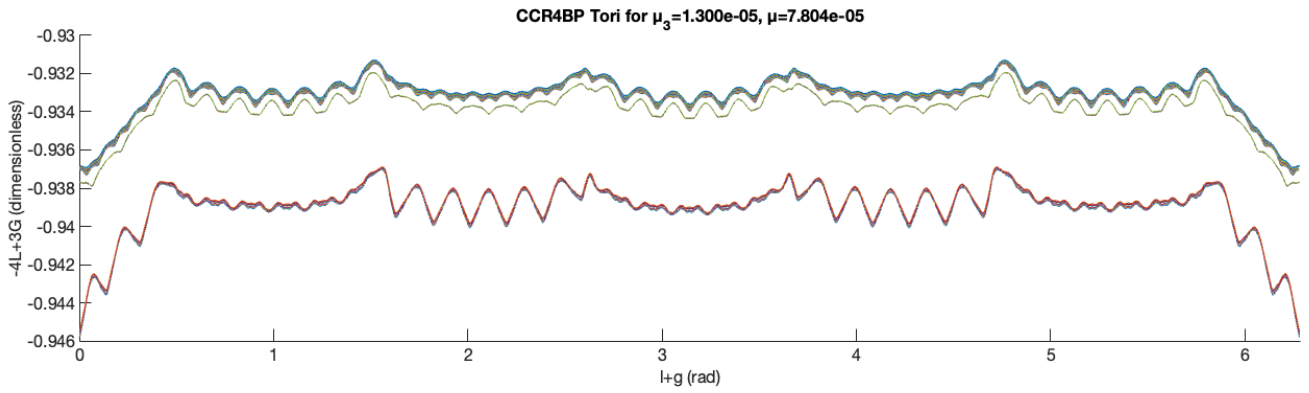}
\includegraphics[width=\columnwidth]{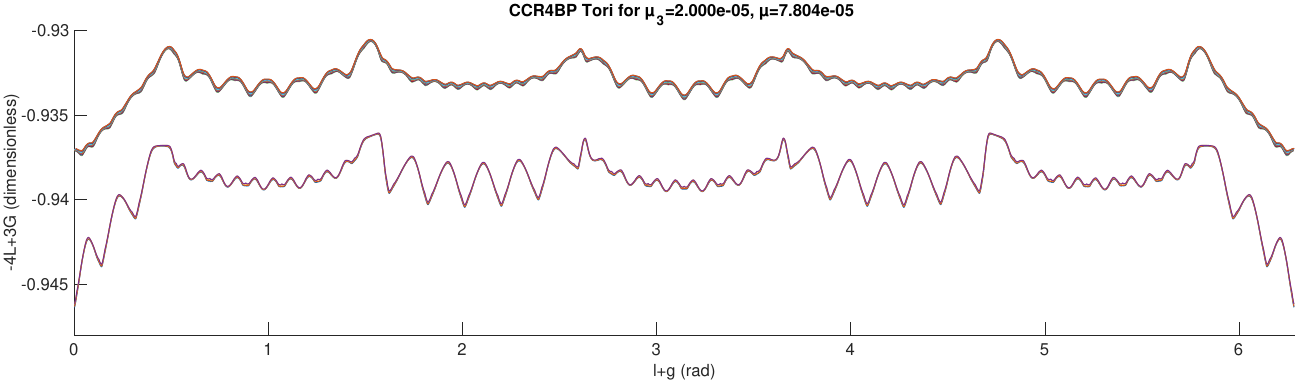}
\caption{Persisting CCR4BP 4:3 J-G orbit tori for $\omega \in [2.032685, 2.0405]$ plotted in $(-Q_{2}, P_{2})$ at $\mu_{3} \times 10^{6}=10.5, 11, 12, 13, 20$ (from top to bottom) }
\label{fig::continuations2} \vspace{-18pt}
\end{centering}
\end{figure}

The results of the continuations of the tori with $\omega \in [2.032685, 2.0405]$ are displayed in Figures \ref{fig::continuations1}-\ref{fig::continuations2}, for the $\mu_{3,j}$ values listed in the figure captions. 
The tori higher on the vertical axis have larger $\omega$, with $\omega$ monotonically decreasing with $P_{2}$. 
Although all the 4:3 J-G orbits inside $\Xi_{\mu_{3}=0}$ were very simple, near-horizontal circulating invariant tori for the J-G PCRTBP, as soon as $\mu_{3}>0$ we start seeing secondary resonance islands appear. For even just $\mu_{3}=1 \times 10^{-6} \approx 0.04 \mu_{3,e}$, very prominent islands are generated by the previously-mentioned secondary resonances $\frac{\omega}{2\pi} = \frac{12}{37}$ and $\frac{11}{34}$; $\frac{12}{37}$ is the large island at the top of that plot, while $\frac{11}{34}$ is the large island at its bottom. However, noticeable islands also appear at higher order secondary resonances $\frac{\omega}{2\pi} = \frac{23}{71}$ and $\frac{34}{105}$, which are expected\cite{morbyBook} to be weaker (for $\frac{\omega}{2\pi} = \frac{m}{n}$, the resonance order is defined as $m+n$).

Increasing $\mu_{3}$ from $1 \times 10^{-6}$ to $3 \times 10^{-6}$ and then $5 \times 10^{-6}$, the previously-seen islands at 12/37, 23/71, 34/105, and 11/34 grow significantly wider, with the tori which formed the bottom edge of the 11/34 island disappearing. However, yet more higher-order secondary resonant islands appear, at $\frac{\omega}{2\pi} = \frac{25}{77}$ and $\frac{35}{108}$. By $\mu_{3}=8 \times 10^{-6} \approx 0.317\mu_{3,e}$, another visibly wide island appears at $\frac{\omega}{2\pi} = \frac{37}{114}$. As is expected, for each secondary resonance $m/n$, there are $n$ pendulum-shaped regions appearing side by side within its island. All of these secondary resonances have order much larger than those generally seen in the context of mean motion resonances (where both the numerator and denominator are usually less than 10), but nevertheless their effects are quite prominent, as evidenced by the growing widths of their corresponding islands. 

Increasing $\mu_{3}$ further until $1 \times 10^{-5}$, we see that the strips of tori separating the large upper 12/37 island from the small 35/108 and 37/114 islands (just below and above 12/37, respectively) have grown very thin. The strips of tori separating other consecutive islands ($\frac{25}{77}, \frac{37}{114}, \frac{12}{37}, \frac{35}{108},  \frac{23}{71},  \frac{34}{105}$, and $\frac{11}{34}$, in that order from top to bottom) have also grown much thinner than for smaller $\mu_{3}$ values. As $\mu_{3}$ grows to $1.05, 1.1, 1.2, 1.3$, and $2 \times 10^{-5}$, these strips of tori separating consecutive secondary resonant islands grow thinner and thinner, and then disappear. Apart from a few of the lowest-energy tori near $\omega=2.0405$, the last circulating invariant circle in this part of the 4:3 J-G family NHIM $\Xi_{\mu_{3}}$ fails to continue past $\mu_{3} = 2.05 \times 10^{-5}$; its $\frac{\omega}{2\pi}$ had a continued fraction expansion of $[0,3,11,3,1,1,3,\bar 1]$, which was one of the ``most noble'' $\omega_{k,S}$ values we started from (in the sense of having the smallest continued fraction terms before $\bar 1$). 

\begin{figure}
    \begin{centering}
\includegraphics[width=0.5\columnwidth]{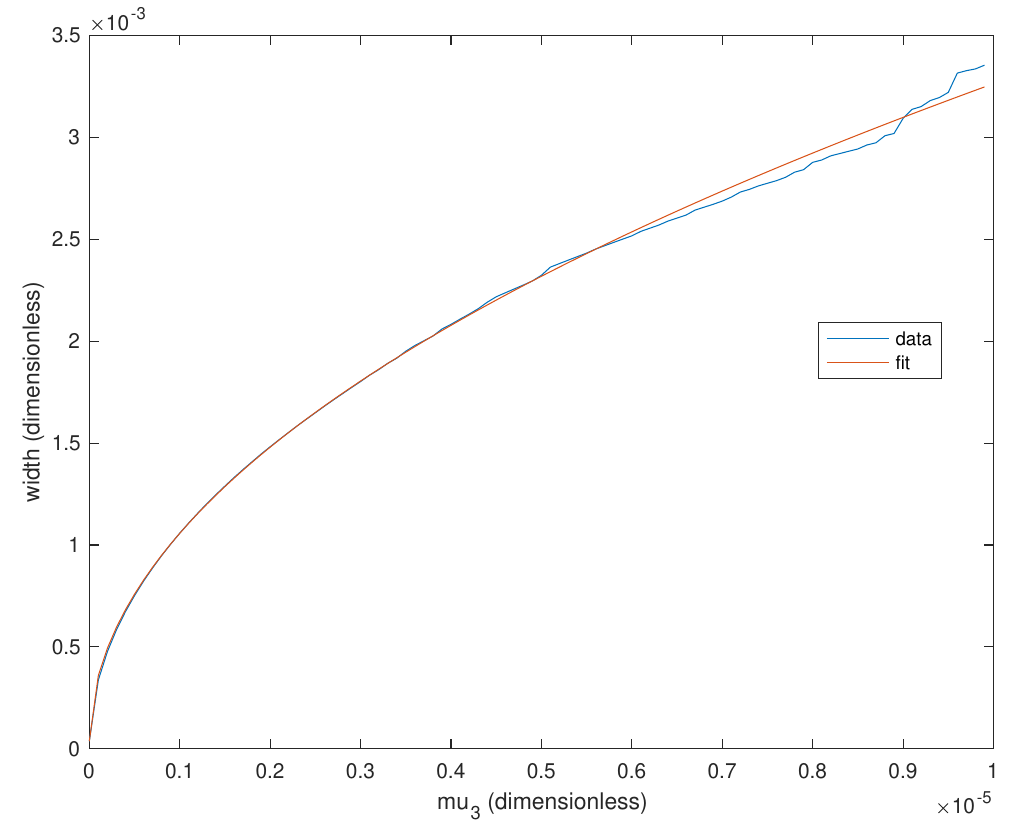}
\caption{12:37 secondary resonance width vs $\mu_{3}$ along with $1.02\sqrt{\mu_{3}}$ fit} \label{fig::width} \vspace{-18pt}
\end{centering}
\end{figure}

To check if there were any secondary resonant $\omega$ which did not have a noticeable effect on these unstable 4:3 J-G tori, we used a Farey sequence\cite{weisstein2006farey} to help find all rational rotation numbers with denominator $<125$ lying in the range $[2.032685, 2.0405]$. The numbers found were $2\pi$ times $\frac{11}{34}, \frac{34}{105}, \frac{23}{71}, \frac{35}{108}, \frac{12}{37}, \frac{37}{114}$, and $\frac{25}{77}$, all of which did generate significant secondary resonant islands. For smaller $\mu_{3}$, we can measure the width of the islands as a function of $\mu_{3}$. This is shown in Figure \ref{fig::width} 
for the 12/37 secondary resonance, where the width is measured as the vertical $P_{2}$ distance at $-Q_{2} = 2\pi \frac{19}{37}$ from the top edge of the island to the bottom edge of the island. As predicted by perturbation theory\cite{morbyBook}, the width scales proportionally with $\sqrt{\mu_{3}}$ for smaller $\mu_{3}$ (here, $1.02 \sqrt{\mu_{3}}$ provides an excellent fit to the data), further confirming the validity of our methodology. 

In summary, the tori disappear first near lower-order secondary resonances, followed by near higher-order ones, as expected. The secondary resonant islands grow wider and wider until no tori remain to separate them, most with widths initially scaling with $\sqrt \mu_{3}$, also as expected. All of the previous evidence thus suggests that the widths of the islands seen in Figures \ref{fig::continuations1}-\ref{fig::continuations2} are correct. The widths are clearly significant, and they also grow fast enough with $\mu_{3}$ that they could overlap. This would then provide the mechanism for the disappearance of tori separating islands (as explained in the section on Chirikov overlapping). Thus, these continuation-based plots very strongly suggest that the secondary resonances overlap inside this higher-energy portion of $\Xi_{\mu_{3}}$ by the time $\mu_{3}$ reaches $\mu_{3,e}$. However, for final confirmation, we need to compute separatrices. 

\section{Final Confirmation of 4:3 J-G Family Secondary Resonance Overlap in the Physical Mass J-E-G through Separatrix Computation} 

As mentioned in the section on the Chirikov criterion, resonant islands overlap in 2D maps when their stable and unstable manifold separatrices intersect. Assuming the 2D cylindrical NHIM $\Xi_{\mu_{3}}$ of unstable 4:3 J-G orbits persists until $\mu_{3}=\mu_{3,e}$, so that we can restrict attention to $\Xi_{\mu_{3}}$ and consider $F_{\mu_{3}}$ as a 2D map, this means that we should check if the separatrices of its secondary resonances intersect. If they do intersect, then this would prove that the circulating invariant circles with $\omega \in [2.032685, 2.0405]$, continued from the $\mu_{3}=0$ PCRTBP in the previous section, cannot persist into the $\mu_{3}=\mu_{3,e}$ physical mass J-E-G CCR4BP. 

\subsection{NHIM-unstable periodic orbit computation and NHIM persistence} 

Recall from the section on the internal structure of $\Xi_{\mu_{3}}$ that each separatrix will emanate from a NHIM-unstable secondary resonant $F_{\mu_{3}}$-periodic orbit. Thus, first these periodic orbits must be computed in the higher-energy region of interest, followed by their stable and unstable manifold separatrices. For a secondary resonance at $\frac{\omega}{2\pi}=\frac{m}{n}$, $m,n \in \mathbb{Z}$ coprime, the period of its periodic orbit will be $n$ iterations of $F_{\mu_{3}}$, with $n$ \emph{discrete} points $X(k)$ lying on that orbit. However, for the $\mu_{3}=0$ PCRTBP, even secondary resonant $\omega$ have corresponding invariant circles, which are continuous objects, not discrete. This means that the $\mu_{3}=0$ invariant circle at a rational $\omega$ must contain a (infinite) continuum of $F_{\mu_{3}=0}$-periodic orbits, all of whose points taken together form that invariant circle. As soon as $\mu_{3}>0$, all of those infinitely many $F_{\mu_{3}=0}$-periodic orbits disappear except two, which persist as the NHIM-stable and NHIM-unstable periodic orbits. 

The continuation from $\mu_{3}=0$ to $\mu_{3}=\mu_{3,e}$ of a NHIM-unstable or NHIM-stable secondary resonant periodic orbit needs to be started from the correct $F_{\mu_{3}=0}$-periodic orbit. To find which two of the infinite continuum of such orbits persists for $\mu_{3}>0$, note that each plot of Figures $\ref{fig::continuations1}$-\ref{fig::continuations2} is symmetric about the line $l+g=\pi$. The symmetry necessitates that for each secondary resonance, either the NHIM-stable or NHIM-unstable periodic orbit must pass through that line so that its row of pendulums is symmetric. Thus, for $\frac{\omega}{2\pi}=\frac{m}{n}$, one of the two persisting secondary resonant $F_{\mu_{3}=0}$-periodic orbits will be the one containing a point $\bold{x}_{0}$ on $l+g=\pi$. The other one will contain the point found by integrating $\bold{x}_{0}$ by the PCRTBP flow for time $\frac{T}{2n}$ ($T$ being that orbit's PCRTBP flow period). The CCR4BP flow has a time-reversal symmetry about the $x$-axis similar to the PCRTBP, which is valid for our stroboscopic map as well due to the earlier choice of stroboscopic map phase $\theta_{p,f}=0$. For each 4:3 J-G $\mu_{3}=0$ circle, $l+g=\pi$ corresponds in cartesian space to the circle's vertical crossing with the positive $x$-axis, which causes the symmetry about $l+g=\pi$. 

\begin{figure}
    \begin{centering}
\includegraphics[width=\columnwidth]{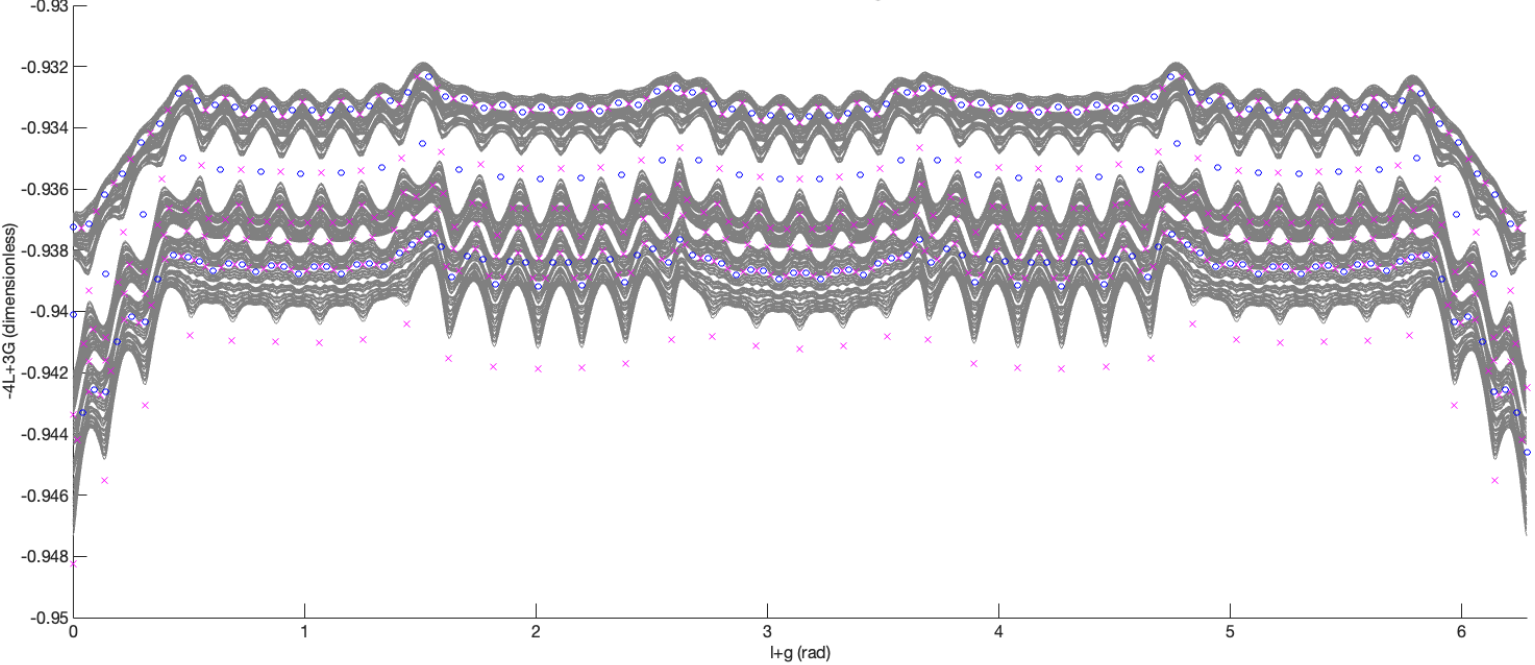}
\caption{Persisting CCR4BP 4:3 J-G tori and secondary resonant periodic orbits for $\mu_{3}=8 \times 10^{-6}$ (magenta x for unstable, blue o for stable, tori in gray for context)} \label{fig::secondaryPos8}
\vspace{-12pt}
\end{centering}
\end{figure}

\begin{figure}
    \begin{centering}
    \includegraphics[width=\columnwidth]{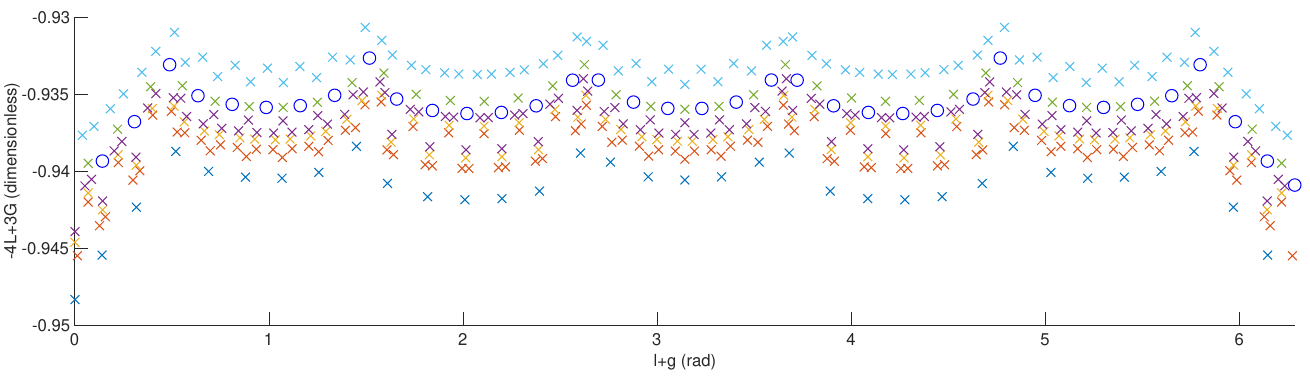}
\caption{$\mu_{3}=\mu_{3, e}$ CCR4BP 4:3 J-G secondary resonant periodic orbits (x for unstable with different colors for different orbits, {blue} {o} for stable)}
\label{fig::secondaryPos}
\vspace{-18pt}
\end{centering}
\end{figure}

\begin{figure}
    \begin{centering}
\includegraphics[width=0.49\columnwidth]{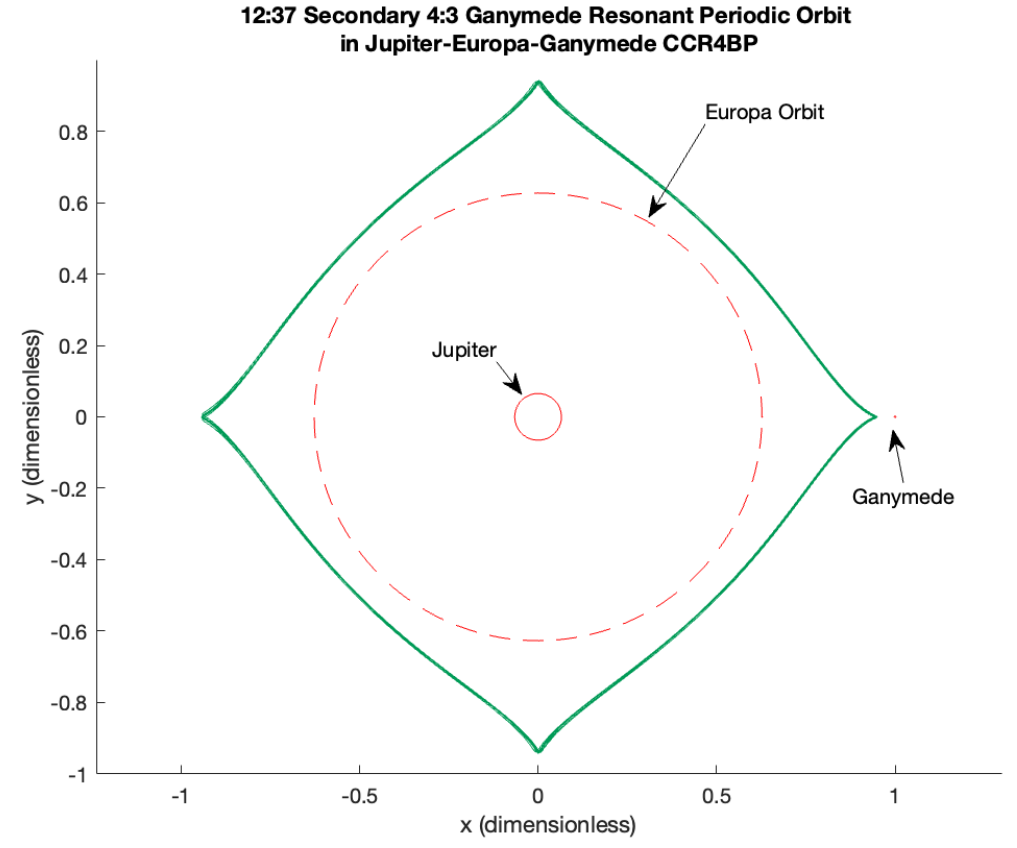}
\includegraphics[width=0.49\columnwidth]{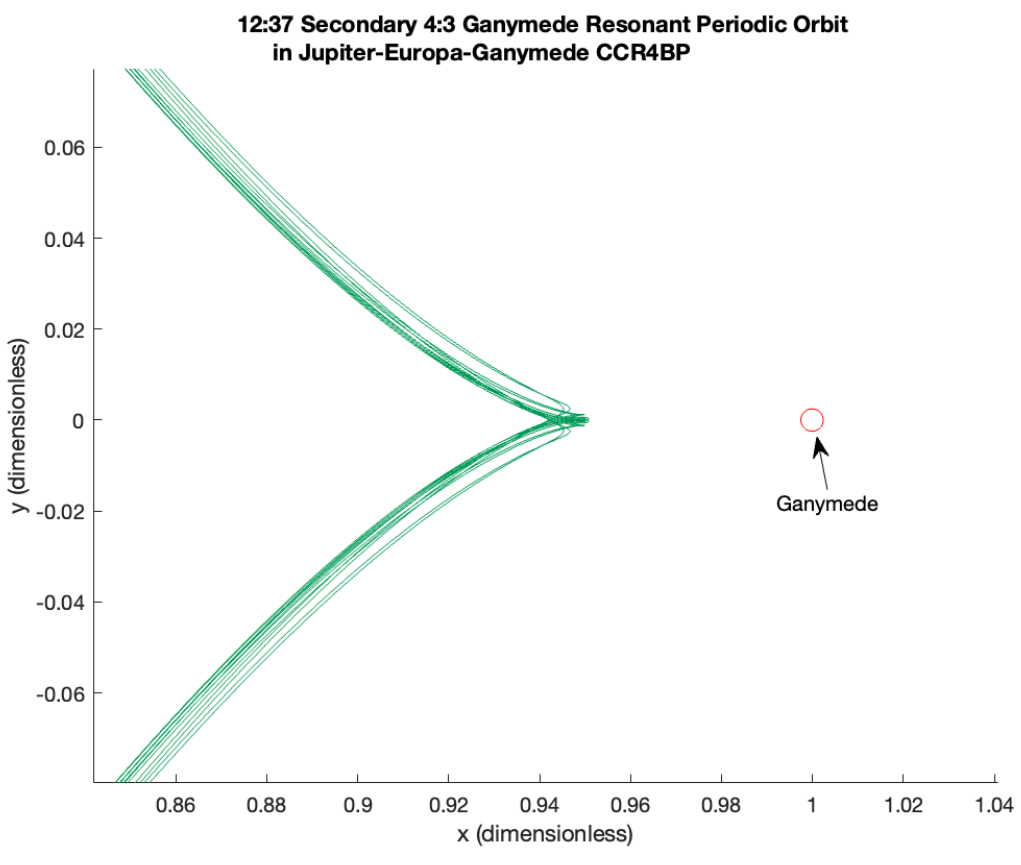}
\caption{12:37 Secondary Resonant 4:3 Ganymede Orbit under $\mu_{3}=\mu_{3,e}$ J-E-G CCR4BP flow} \label{fig::1237po}
\vspace{-12pt}
\end{centering}
\end{figure}

As found in the previous section, the secondary resonances $\frac{11}{34}, \frac{34}{105}, \frac{23}{71}, \frac{35}{108}, \frac{12}{37}, \frac{37}{114}$, and $\frac{25}{77}$ were the most important. Using the previous discussion on persisting $F_{\mu_{3}}$-periodic orbit locations, we used the multiple-shooting quasi-Newton method described earlier to successfully continue all of those NHIM-unstable secondary resonant periodic orbits from $\mu_{3}=0$ to the physical $\mu_{3,e}$, except for $\frac{37}{114}$ due to a NHIM-stable to NHIM-unstable bifurcation (we anticipate computing this last one in the near future as well, after some small extensions to our computer implementation required to handle complex Floquet multipliers). For visualization of how and where secondary resonant periodic orbits appear inside islands, in Figure \ref{fig::secondaryPos8} we plot both the persisting invariant circles as well as the secondary resonant periodic orbit points for the $\mu_{3}=8 \times 10^{-6}$ CCR4BP. As expected from the pendulum phase portrait of Figure \ref{fig::pendulum}, NHIM-unstable orbit points occur at the ``necks'' between pendulums, while NHIM-stable ones occur at the centers of their ``bulges''. We also plot all the secondary resonant periodic orbit points for $\mu_{3}=\mu_{3,e}$ in Figure \ref{fig::secondaryPos}, as well as a single example secondary resonant periodic orbit under the continuous-time J-E-G CCR4BP flow in Figure \ref{fig::1237po}. 

Each NHIM-unstable $F_{\mu_{3}}$-periodic orbit has two stable and two unstable Floquet multipliers, all of which are found as part of the multiple-shooting quasi-Newton method. One stable/unstable multiplier pair $\lambda_{s}= \lambda_{u}^{-1}$ corresponds to the stable/unstable directions \emph{transverse} (not tangent) to the 4:3 J-G orbit NHIM $\Xi_{\mu_{3}}$, while the other pair $\lambda_{1}= \lambda_{2}^{-1}$ will be for the stable/unstable directions tangent to $\Xi_{\mu_{3}}$ generated by its secondary resonance island (assuming that $\Xi_{\mu_{3}}$ persists). Recall from the section on NHIMs that the condition defining a NHIM is that it has transverse directions contracting/expanding more strongly than its tangent directions; furthermore, this condition is what is also required for it to persist under perturbations such as, in our case, $\mu_{3} > 0$. Thus, for the NHIM-unstable orbits computed in the $\mu_{3}=\mu_{3,e}$ physical J-E-G CCR4BP, we can compare the tangent Floquet multiplier pair $(\lambda_{1}, \lambda_{2})$ with the transverse pair $(\lambda_{s}, \lambda_{u})$. What we found is that indeed there remains a large difference in the tangent versus transverse contraction/expansion rates. The strongest tangent $\lambda_{2}$ was $\approx 1.035$, while the weakest transverse $\lambda_{u}$ among these higher-energy orbits was $\approx  2.508$. This indicates that the NHIM $\Xi_{\mu_{3}}$ indeed persists until $\mu_{3}=\mu_{3,e}$. 

\vspace{-2pt}
\subsection{Separatrix computation and confirmation of secondary resonance overlap} 
\vspace{-2pt}

\begin{figure}
    \begin{centering}
    \includegraphics[width=\columnwidth]{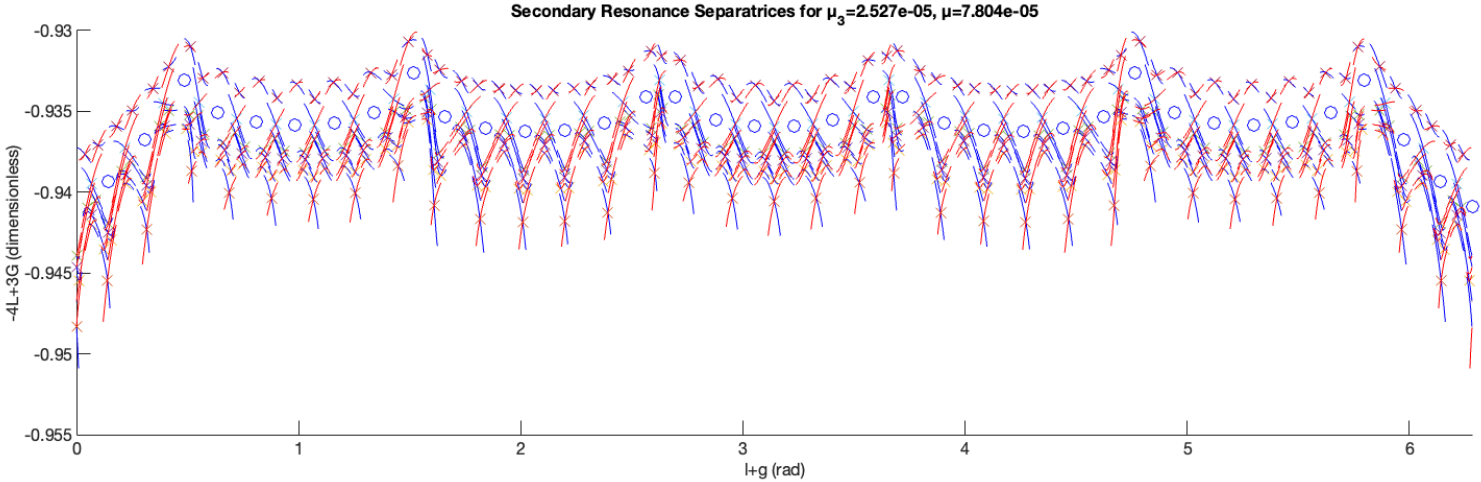}
\caption{$\mu_{3}=\mu_{3, e}$ CCR4BP 4:3 J-G separatrices ({red} for unstable, {blue} for stable), plus secondary resonant periodic orbits (various color x for unstable, {blue} {o} for stable)}
\label{fig::separatrices}
\vspace{-16pt}
\end{centering}
\end{figure}

With the NHIM-unstable secondary resonant periodic orbits continued into the physical J-E-G $\mu_{3}=\mu_{3,e}$, we  used the previously-described stable/unstable manifold separatrix parameterization method to finally compute their attached separatrices. As mentioned in the computations section, each separatrix is represented by a set of $n$ polynomials with some finite domain of validity; plotting the curves parameterized by these polynomials over their domains of validity in $(-Q_{2}, P_{2})$ space, we get Figure \ref{fig::separatrices}. Unstable manifold separatrices are shown in red, while stable ones are in blue. Due to the symmetry described in the previous section, the unstable separatrices are just the reflections of the stable ones about the line $\ell + g=\pi$. As discussed in the section on computing these parameterizations, we do not use numerical integration to globalize these stable/unstable manifold separatrix curves, as the strong expansion in directions transverse to the NHIM $\Xi_{\mu_{3}}$ would amplify any errors and push the resulting curves off of $\Xi_{\mu_{3}}$. Nevertheless, even without globalization, we are able to observe intersections between the separatrices of different secondary resonances. 

    \begin{figure}
    \begin{centering}
    \includegraphics[width=\columnwidth]{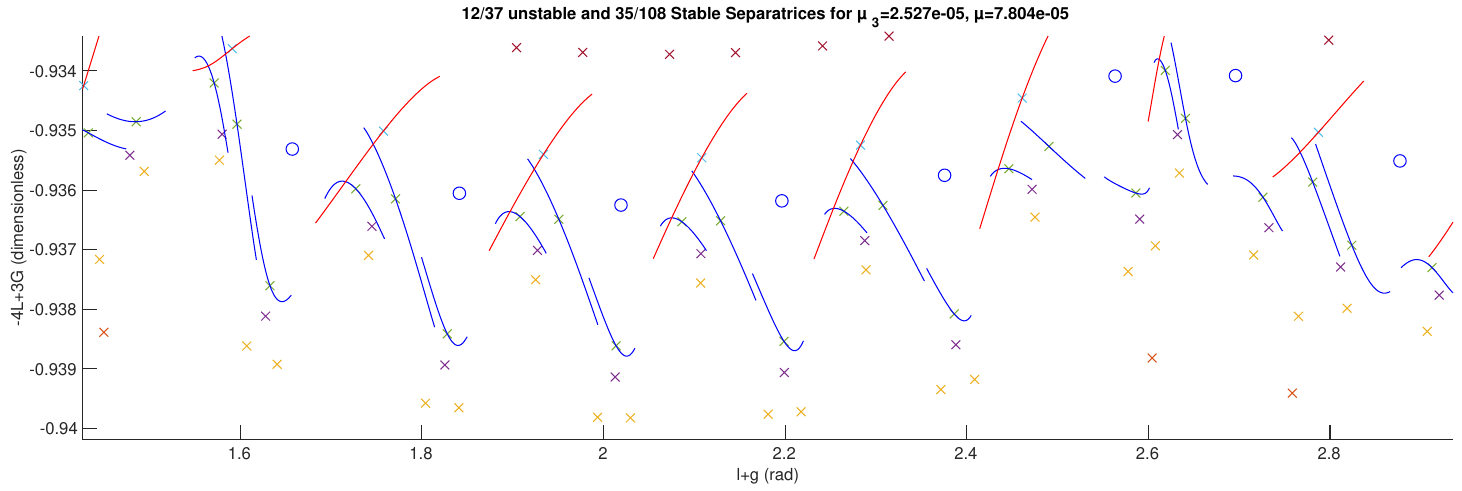}
\caption{$\mu_{3}=\mu_{3, e}$ CCR4BP 4:3 J-G 12/37 unstable (red) and 35/108 stable (blue) separatrices}
\label{fig::separatricesIntersect}
\vspace{-20pt}
\end{centering}
\end{figure} 

Since it is difficult in Figure \ref{fig::separatrices} to distinguish separatrices of different secondary resonances from each other, we can instead plot the unstable separatrix for each island and the stable separatrix of its adjacent island two at a time to more easily observe intersections between them. Recall that the successfully computed secondary resonant NHIM-unstable periodic orbits (and thus separatrices) were at $\frac{11}{34}, \frac{34}{105}, \frac{23}{71}, \frac{35}{108}, \frac{12}{37}$, and $\frac{25}{77}$, in that order. For each plotted pair of consecutive secondary resonances $\frac{11}{34}, \frac{34}{105}, \frac{23}{71}, \frac{35}{108}$, and $ \frac{12}{37}$, such intersections of unstable and stable separatrices were indeed observed. An example plot showing such intersections for the stable $\frac{35}{108}$ and unstable $ \frac{12}{37}$ separatrices is shown in Figure \ref{fig::separatricesIntersect}; due to the symmetry around $l+g=\pi$, reversing the choice of stable and unstable separatrices will also yield the same plot and separatrix intersections, except with a reflection. No intersection between the separatrices of $\frac{12}{37}$ and $\frac{25}{77}$ was detected at the level of manifold globalization seen in Figure \ref{fig::separatrices}, but recall that the secondary resonant orbit at $\frac{37}{114}$ lying between $\frac{12}{37}$ and $\frac{25}{77}$ is still yet to be computed. We believe that once this last periodic orbit and its separatrices are computed, we will find separatrix intersections in this orbit family region as well. 

As a final check on the persistence of the unstable 4:3 J-G orbit family NHIM $\Xi_{\mu_{3}}$ until $\mu_{3}=\mu_{3,e}$, we can also check whether intersections of separatrix curves in the 2D plot of Figure \ref{fig::separatrices} are also intersections of those curves in the full $F_{\mu_{3}}$ phase space $\mathbb{R}^{4}$. Checking this for a number of such 2D plot intersections, the previous condition was indeed satisfied. However, 1D curves generically do not intersect in 4D space; they do only when contained inside a 2D manifold. Thus, in combination with the tangent vs transverse expansion/contraction rates discussed in the previous section, this provides extremely strong evidence that the 2D cylindrical NHIM $\Xi_{\mu_{3}}$ persists until $\mu_{3}=\mu_{3,e}$ and contains all of these separatrices. Given the separatrix intersections inside $\Xi_{\mu_{3}}$ shown in Figures \ref{fig::separatrices}-\ref{fig::separatricesIntersect}, we hence have full, final confirmation that secondary resonances $\frac{11}{34}, \frac{34}{105}, \frac{23}{71}, \frac{35}{108}$, and $ \frac{12}{37}$ overlap in the NHIM $\Xi_{\mu_{3}}$ for $\mu_{3}=\mu_{3,e}$. 

The overlapping of these secondary resonances also confirms that the higher-energy unstable 4:3 J-G circulating invariant circles with $\omega \in [2.032685, 2\pi\frac{12}{37} \approx 2.037790]$, which exist in the $\mu_{3}=0$ PCRTBP case, are destroyed before the $\mu_{3}=\mu_{3,e}$ physical mass J-E-G CCR4BP. We expect to get the same result for the remaining orbits, for $\omega \in [2.037790, 2.0405]$, once the 37/114 secondary resonant periodic orbit and separatrices are found as well. Nevertheless, though these invariant circles corresponding to PCRTBP unstable 4:3 J-G periodic orbits cannot be continued into, and thus have no dynamical equivalent in, the J-E-G CCR4BP, the previous discussion shows that the \emph{family} of these orbits represented by the NHIM $\Xi_{\mu_{3}}$ \emph{does} have a dynamical equivalent in the J-E-G CCR4BP. The key is that at higher energies, the dynamics \emph{inside} the NHIM change completely, from one dominated by circulating invariant circles in the J-G PCRTBP to one governed by secondary resonances and their NHIM-unstable periodic orbits and separatrices in the J-E-G CCR4BP. 

\section{Conclusions}

In this study, we showed that the effect of Europa on the unstable Jupiter-Ganymede 4:3 mean motion resonant orbits causes a major qualitative change in the structure of that orbit family, as compared to the PCRTBP. This occurs through the generation of secondary resonances between the PCRTBP orbit periods and the perturbation of Europa, and the overlap of these secondary resonances inside of the normally hyperbolic invariant manifold formed by the 4:3 unstable orbits. Through continuations of invariant circles, followed by computations of secondary resonant periodic orbits and their attached stable/unstable manifold separatrices, we not only confirmed the secondary resonance overlap and the resulting destruction of dynamical equivalents of the higher-energy PCRTBP periodic orbits, but also found that the orbit family as a whole does persist into the physical-mass Jupiter-Europa-Ganymede CCR4BP. New types of orbits, different topologically from the PCRTBP ones, appear instead inside the family at those higher energies. This overlap thus has major implications for the types of resonant orbits which should be analyzed in certain multi-moon systems.

Although this study was done for the Jupiter-Ganymede 4:3 resonant orbit case, we believe that this same behavior will occur in many other families of unstable mean motion resonant orbits as well. In fact, the failure of higher-energy orbits to continue until $\mu_{3}=\mu_{3,e}$, described earlier in this paper for the 4:3 orbits, also occurred in our previous Jupiter-Europa-Ganymede CCR4BP based study\cite{kumar2023} for the unstable Jupiter-Ganymede 7:5 and 3:2 mean motion resonant orbits. Secondary resonance overlap would explain the continuation failure for those MMRs as well. Furthermore, such phenomena likely occur at high energies and eccentricities inside MMRs with bodies other than Ganymede as well, as long as a strong enough perturbing body exists. For instance, a higher-$a$ Europa exterior unstable MMR orbit could experience secondary resonances with Ganymede; or in the Uranian system, an Oberon interior unstable MMR orbit could be influenced by Titania in a similar manner. All of these moons have mass ratios on the order of $10^{-5}$ with their planets, so the perturbations would be of a similar order of magnitude to the one studied in this paper. Thus, there are many systems and MMRs for which an investigation of secondary resonant phenomena should be done. 

For investigation of unstable mean motion resonant orbits in a CCR4BP model, it should be kept in mind that secondary resonances \emph{may} play a decisive role, but that there are also cases where this does not happen; for example, the Jupiter-Europa 3:4 PCRTBP unstable resonant periodic orbit family has previously\cite{kumar2021aug} been successfully continued into the Jupiter-Europa-Ganymede CCR4BP as invariant tori across a wide range of energies and $\lambda_{u}$. Thus, for investigating the effect of adding a third large body on an unstable PCRTBP resonant orbit family, we suggest first attempting to continue the orbits as tori into the corresponding CCR4BP. If these continuations fail for some range of $\omega$ at higher energies, then a Farey sequence can be used to identify all secondary resonant rotation numbers below a certain order in that $\omega$ range; these will be the secondary resonances whose NHIM-unstable periodic orbits and separatrices should be computed. 

As mentioned in the introduction, the primary importance of mean motion resonances is that overlapping of different MMRs (i.e. heteroclinics between their unstable orbits) drives changes of semi-major axis, and thus large scale transport across the phase space. With the knowledge gained in this study, we now know what types of unstable resonant orbits can exist at higher energies in higher-fidelity models like R4BPs. Crucially, such high-energy orbits make close flybys of their resonant moon and thus have higher $\lambda_{u}$-values, so their stable/unstable manifolds will yield transfers with lower TOF - critical for practical missions. They are also more likely to have heteroclinics with other unstable MMR orbits. With the resonant orbits found, we can now compute these manifolds. In our previous work\cite{kumar2023}, we searched for transfers from the Jupiter-Ganymede 4:3 MMR to the Jupiter-Europa 3:4, but were unable to find heteroclinics due to the lack of high-$\lambda_{u}$ 4:3 orbits found. Now that those orbits have been characterized and in many cases (periodic orbits and separatrices) computed, we can resume this heteroclinic search, which showed promise even in that earlier study. 

Since the high-energy, high-$\lambda_{u}$ 4:3 Jupiter-Ganymede unstable resonant orbits occur inside secondary resonances, they will include NHIM-unstable periodic orbits, whose stable/unstable manifolds will have to be computed. These will be 2D in the entire CCR4BP phase space, as opposed to 1D when considered only inside the 2D NHIM of orbits; in fact, the 1D separatrices of each secondary resonant periodic orbit will be contained inside the 2D stable/unstable manifolds of that same orbit. A parameterization method for 2D stable/unstable manifolds of map-periodic orbits, similar to those\cite{haroetal} for 2D stable/unstable manifolds of fixed points, should be possible to develop for this. The librational tori inside secondary resonances can also be computed, but will require an extension of our torus computation method\cite{kumar2022} to multiple-shooting. We are confident that such secondary resonant orbits inside MMRs and their stable/unstable manifolds, both in the Jovian system as well as others, will provide useful transfer options for missions in multi-moon systems. 
\vspace{-1pt}
\subsection{Acknowledgements}

This work was supported by the National Science Foundation under award no. DMS-2202994. This research was carried out at the Jet Propulsion Laboratory, California Institute of Technology, under a contract with the National Aeronautics and Space Administration (80NM0018D0004). The HPC resources used in this investigation were provided by funding from the JPL Information and Technology Solutions Directorate. RLA was supported through funding by the Multimission Ground System and Services Office (MGSS) in support of the development of the Advanced Multi-Mission Operations System (AMMOS).

\vspace{-2pt}

\bibliographystyle{AAS_publication}   
\bibliography{references}   

\end{document}